%% file: main.tex
\documentclass[sigconf,9pt,screen]{acmart}

\AtBeginDocument{
  }

\renewcommand\footnotetextcopyrightpermission[1]{}

\usepackage{tikz}
\usepackage{braket}
\usepackage{lipsum}
\usepackage{subfig}
\usepackage{xcolor}
\usepackage{amsmath}
\usepackage{balance}
\usepackage{environ}
\usepackage{physics}
\usepackage{wrapfig}
\usepackage{colortbl}
\usepackage{enumitem}
\usepackage{amsfonts}
\usepackage{booktabs}
\usepackage{graphicx}
\usepackage{textcomp}
\usepackage{adjustbox}
\usepackage{algorithm}
\usepackage{algpseudocode}
\usepackage{arydshln}
\usepackage[normalem]{ulem}

\newcommand{\sol}{SpinTune}

\begin{document}

\copyrightyear{2026}
\acmYear{2026}
\setcctype{}
\acmConference[]{}
\acmBooktitle{}
\acmDOI{}
\acmISBN{}

\title[\sol{}: Improving the Reliability of Quantum Sensor Networks for Practical Quantum-Classical Utility]{\sol{}: Improving the Reliability of Quantum Sensor Networks for Practical Quantum-Classical Utility}

\author{Jason Ludmir}
\affiliation{%
  \institution{Rice University}
  \city{Houston}
  \state{TX}
  \country{USA}
}
\email{jzl2@rice.edu}

\author{Nicholas S. DiBrita}
\affiliation{%
  \institution{Rice University}
  \city{Houston}
  \state{TX}
  \country{USA}
}
\email{np52@rice.edu}

\author{Jason Han}
\affiliation{%
  \institution{Rice University}
  \city{Houston}
  \state{TX}
  \country{USA}
}
\email{jh146@rice.edu}

\author{Tirthak Patel}
\affiliation{%
  \institution{Rice University}
  \city{Houston}
  \state{TX}
  \country{USA}
}
\email{tp53@rice.edu}


\begin{abstract}

Emerging quantum sensors are increasingly envisioned as components of hybrid quantum-classical high-performance computing, enabling new capabilities in scientific, cyber-physical, and machine-learning pipelines. However, their practical utility is limited by environmental decoherence, which degrades sensing reliability. While dynamical decoupling (DD) pulse sequences can mitigate this, standard methods are often suboptimal in the presence of realistic noise. We present \sol{}, a reinforcement learning software approach that autonomously discovers adaptive, piecewise DD sequences tailored to specific environments. Using a simulation model of a Carbon-13 spin bath, we show that \sol{} significantly outperforms standard DD sequences in preserving coherence.

We also derive a metric connecting coherence to AC magnetometry sensitivity, showing that \sol{} improves average sensitivity by over 80\% compared to the next-best approach. A case study of a real neutral-atom system confirms these gains, positioning \sol{} as a scalable control layer for reliable quantum sensing in QML pipelines and networked sensing applications.

\end{abstract}

\maketitle

\input{sections/introduction}
\input{sections/background}
\input{sections/motivation}
\input{sections/design}
\input{sections/methodology}
\input{sections/evaluation}
\input{sections/casestudy}
\input{sections/related_work}
\input{sections/conclusion}

\balance

\bibliographystyle{ACM-Reference-Format}
\bibliography{main}

\end{document}

%% file: sections/introduction.tex
\section{Introduction}
\label{sec:introduction}

\begin{figure}[t]
    \centering
    \includegraphics[width=0.99\linewidth]{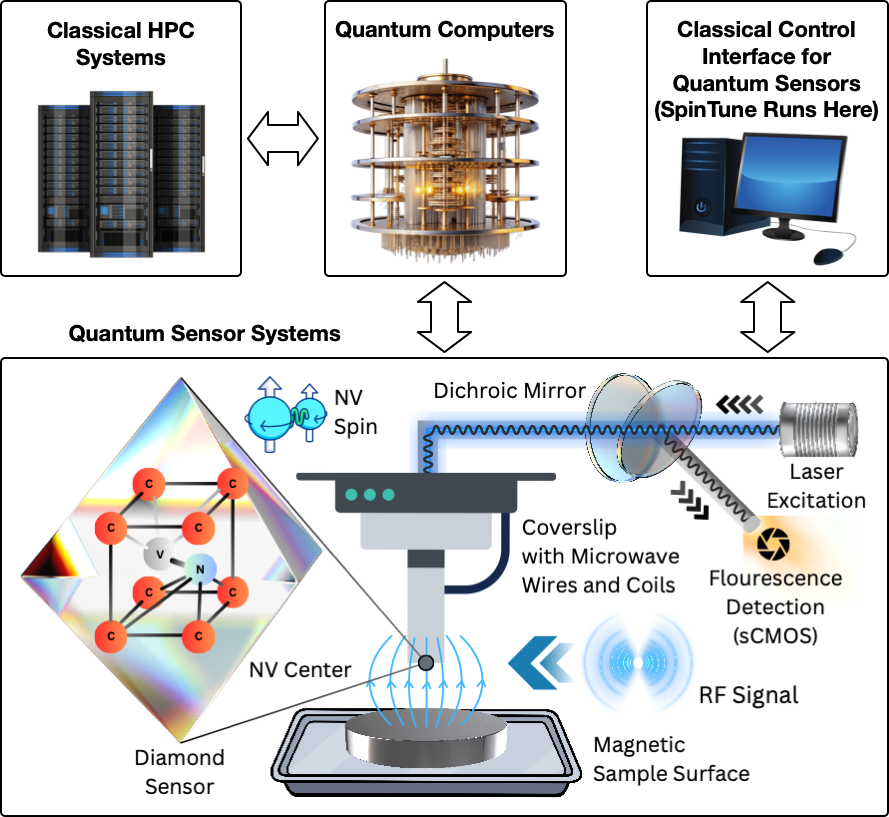}
    \vspace{-3mm}
    \caption{Quantum-classical data and software pipeline and experimental setup to use NV centers as sensor networks for generating quantum data (e.g., AC magnetometers).}
    \vspace{-6mm}
    \label{fig:main_coh}
\end{figure}

\noindent\textbf{Quantum Sensor Systems:} Emerging quantum sensors are increasingly being integrated as components of quantum-classical high-performance computing (HPC) systems, where they act as high-fidelity data sources for cyber-physical, scientific, and machine learning pipelines~\cite{zhang_sensing2023,jiao2024sensing}. An example data flow is depicted in Fig.~\ref{fig:main_coh}, where sensor data is directly fed into quantum computers for quantum machine learning (QML). Recent work demonstrates their deployment in networked settings, including quantum-enhanced WiFi and mmWave sensing infrastructures~\cite{zhang_sensing2023}, as well as fine-grained liquid recognition using distributed quantum sensing~\cite{jiao2024sensing}. Among the most mature platforms are nitrogen-vacancy (NV) centers in diamond and neutral-atom (NA) systems, which offer long coherence times and sensitivity to extremely small magnetic and electrical field variations~\cite{hong2013,takayuki2017,kumar24}. NV centers and NA-based sensors expose quantum states of physical systems, enabling data acquisition with properties such as superposition and entanglement~\cite{Stone_2023,crawford2021,Marciniak_2022,ma21}. Such quantum-generated data has been shown, in theory, to offer advantages for machine learning workloads, including metrology and medical imaging~\cite{mzyk2022,BALASUBRAMANIAN201469,pinilla22}. As quantum sensor networks are increasingly coupled with large-scale ML and scientific pipelines, ensuring reliable, performance-aware operation of these sensors becomes a system-level challenge rather than a purely device-level concern.

\vspace{4mm}

\noindent\textbf{The Challenge:} The practical use of NV or NA-based quantum sensors in scientific computing pipelines is fundamentally limited by decoherence -- the loss of quantum information due to environmental interactions~\cite{Szankowski_2017,hernandez18,aharon2019}. In diamonds, the dominant decoherence source is a fluctuating bath of $^{13}$C (Carbon-13) nuclear spins. These environmental perturbations introduce time-varying magnetic fields that degrade NV spin coherence, limiting sensing quality and reliability. Enhancing NV coherence is therefore critical to advancing quantum sensing toward broader deployment. Dynamical decoupling (DD) techniques -- which apply structured sequences of control pulses to suppress decoherence -- are a powerful approach for prolonging NV coherence~\cite{han2025,lange2010,Viola1999,Yang2011,pershin2025}. Sequences like Carr-Purcell (CPMG) or Uhrig DD (UDD) are known to target specific noise spectral characteristics, effectively acting as frequency-domain filters that quench low-frequency noise~\cite{meiboom1958modified,uhrig2007keeping}. Yet, despite their success, standard DD sequences are often designed analytically under simplified noise assumptions and are not guaranteed to perform optimally in realistic spin environments~\cite{aharon2019,miao2022}. 

\vspace{4mm}

\noindent\textbf{The Opportunity:} In practice, the noise profile experienced by an NV center is highly non-trivial: a complex function of material defects, $^{13}$C spin distributions, and control imperfections~\cite{doherty2013nitrogen,Balasubramanian2009,BarGill2013}. Even under simplified models, the number of possible DD sequence combinations grows exponentially with sequence length, making brute-force optimization infeasible. As shown in recent efforts~\cite{zhang_sensing2023,jiao2024sensing}, quantum sensors are moving into embedded contexts where reliability and adaptivity are as critical as sensitivity. This creates the need for scalable control methods that enable quantum sensors.

\vspace{4mm}

\noindent\textbf{Our Solution:} To tackle this challenge, we propose \sol{}, a first-of-its-kind reinforcement learning (RL)-guided systems framework that automatically discovers high-fidelity, adaptive DD sequences tailored to specific noise environments. \sol{} formulates DD design as a sequential decision-making problem: an agent selects a sequence of pulse segments to maximize final coherence. Unlike classical optimization methods, \sol{} learns by iteratively refining its action policy through interaction with a variety of simulated noise environments that model NV dynamics under a realistic Gaussian noise spectrum. \sol{} introduces several design innovations in this domain, including the use of Fourier transforms to accelerate learning, generalizable model learning applicable to any noise environment, and discretized handling of DD sequences for flexible integration. \textbf{\sol{} is open-sourced at:} \textit{\url{https://github.com/positivetechnologylab/SpinTune}}.

\vspace{4mm}

\noindent\textbf{The contributions of this work include:}

\begin{itemize}[leftmargin=*]

    \item We propose \sol{}, a novel RL framework that constructs piecewise DD sequences tailored to different noise environments.

    \vspace{2mm}
    
    \item \textit{We make three key contributions that depart from prior work.}
    
    \begin{enumerate}
        \item \sol{} embraces a discretized, piecewise representation of DD sequences, enabling efficient RL-based exploration over a combinatorial space of filter functions.
    
        \item It avoids explicit noise spectroscopy or analytical modeling, allowing \sol{} to generalize to environments where spectral properties are only implicitly defined.
        
        \item \sol{} exploits memoization and modular Fourier transforms to substantially accelerate coherence evaluation, making RL feasible at the scale of high-throughput simulation.
    \end{enumerate}

    \vspace{2mm}
    
    \item We build a simulator to extensively evaluate \sol{} against established DD baselines (e.g., UDD, Hahn, CPMG) and a theoretical Oracle across hundreds of noise configurations. Our results show that \sol{} outperforms standard sequences by 35\% and reaches within 18\% of the Oracle performance even after a long evolution duration of $200\,\text{\textmu s}$. \sol{} thus extends NV coherence across a wide range of evolution times.

    \vspace{2mm}

    \item We also derive the relationship between an NV center sensor's coherence under DD sequences and the resulting AC magnetometry sensitivity. We do this by defining a relative sensitivity metric using an NV center controlled by a DD sequence with a fixed sensing time. We show using our derivation that \sol{} improves the average sensitivity of an NV center sensor by over 80\% compared to the next-best baseline DD technique.

    \vspace{2mm}

    \item While primarily focusing on NV centers, we also run a case study of our work on a real neutral-atom testbed, demonstrating 66\% coherence, while the baseline case completely decoheres.

    \vspace{2mm}

    \item \sol{}'s open-sourced implementation is adaptive, general, and scalable, and positions it as a practical tool for generating robust control sequences for NV centers, paving the way for deployable quantum sensors for real-world computing workflows.
\end{itemize}

%% file: sections/background.tex
\section{Relevant Background}
\label{sec:background}

\subsection{Nitrogen-vacancy (NV) centers}

Nitrogen-vacancy centers in diamonds are point defects consisting of a nitrogen substitution adjacent to a lattice vacancy (Fig.~\ref{fig:main_coh}). This corresponds to a grid of carbon atoms in a diamond, where one carbon atom is replaced by a nitrogen atom, and its neighboring position is empty. This ``defect'' in the crystal lattice creates a site with quantum properties (the NV center). The NV center’s electron spin can be initialized and read out optically and coherently controlled with microwave pulses even at room temperature~\cite{lange2010}. These properties, along with long room-temperature spin coherence times, make NV centers a leading platform for quantum sensing and nanoscale magnetometry~\cite{schirhagl2014, aharon2019,BALASUBRAMANIAN201469,hong2013}. In sensing applications, the NV spin serves as an atomic-scale probe of magnetic fields, but its performance is limited by \textbf{decoherence} – the loss of quantum coherence due to interactions with the environment~\cite{maze2013decoherence,Szankowski_2017}. In high-purity diamond, the predominant decoherence mechanism is the coupling between the NV’s electron spin and a ``spin bath'' of $^{13}$C nuclear spins in the diamond lattice: The $^{13}$C atoms' nuclei interact with the NV center’s electron spin in an interfering manner. These environmental spins produce fluctuating local magnetic fields that randomly perturb the NV spin’s phase, causing the NV’s coherent superposition to decay over time.

One method to combat decoherence is \textbf{dynamical decoupling (DD)}, which involves applying a sequence of pulses to the NV spin during its free evolution. DD sequences work by repeatedly flipping the spin’s state, effectively averaging out the phase accumulation from low-frequency noise and filtering the NV’s exposure to environmental fluctuations~\cite{hernandez18}. In essence, the DD pulses act as a filter in the frequency domain: certain noise frequencies are ``quenched'' by the pulse sequence, as the NV center becomes insensitive to noise components slower than the pulse rate~\cite{biercuk2009optimized, cywinski2008enhance}. Classic examples of DD sequences include the Hahn echo (a single $\pi$ pulse)~\cite{hahn1950spin} and multi-pulse sequences like Carr-Purcell-Meiboom-Gill (CPMG) and its variants~\cite{meiboom1958modified}, as well as non-uniform pulse spacing schemes like Uhrig DD (UDD)~\cite{uhrig2007keeping}. Each of these sequence types is known to extend the NV’s spin coherence by suppressing specific spectral components of the noise. By applying a series of pulses, NV coherence times can be prolonged by orders of magnitude~\cite{han2025}. We now explain how to compute the coherence of an NV center under the effects of a particular sequence. 

\subsection{Computing Coherence with Filter Functions}

\begin{figure}[t]
    \centering
    \includegraphics[width=0.99\linewidth]{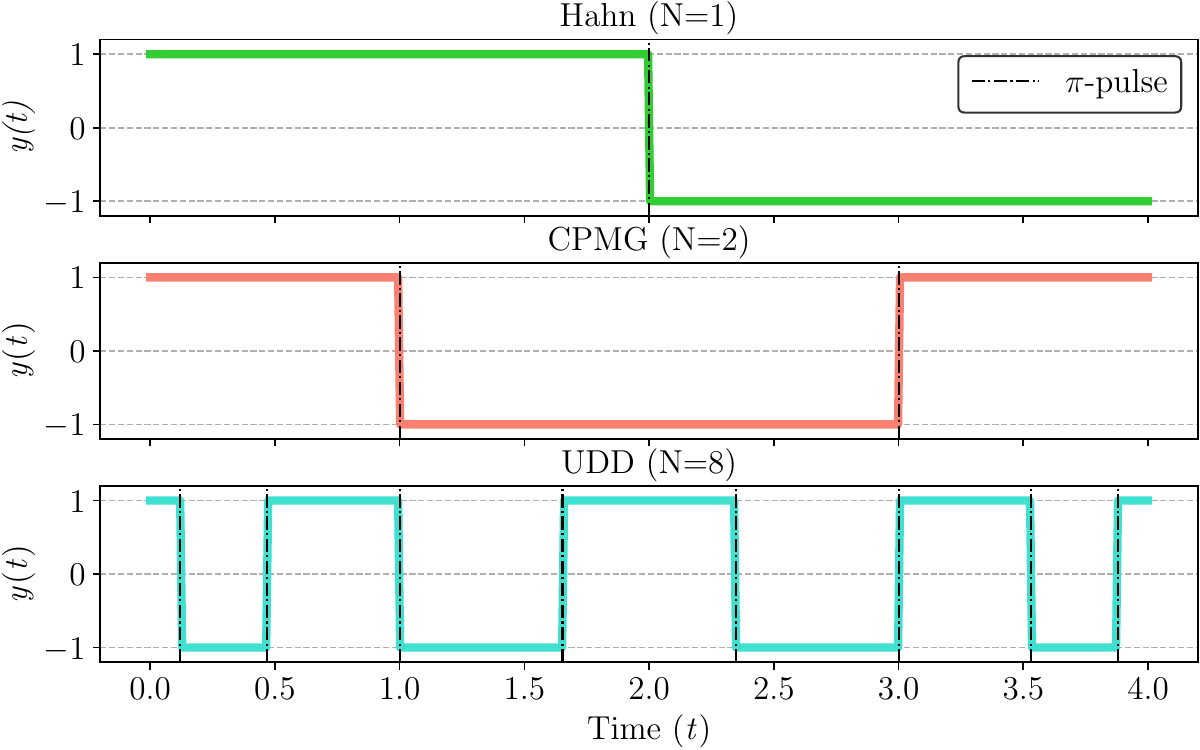}
    \vspace{-3mm}
    \caption{Modulation function $y(t)$ for the standard dynamical decoupling sequences, all shown over a total evolution time $T=4\,\text{\textmu s}$. Note that while we model instantaneous pulse changes for visualization, in real-world settings, microwave pulses for NV centers have some slope.}
    \vspace{-5mm}
    \label{fig:pulse_shapes}
\end{figure} 

To evaluate the effectiveness of a given DD sequence, we calculate the final qubit coherence, $W(T)$, remaining at the total evolution time $T$. Coherence quantifies the preservation of quantum information and is calculated within the widely used filter function formalism~\cite{degen2017,biercuk2009optimized}. In this framework, coherence is given by $W(T) = \exp(-\chi(T))$, where $\chi(T)$ is the \textbf{decoherence function}. It represents the accumulated phase error due to environmental noise and is determined by the overlap between the noise spectrum and the filtering characteristics of the DD sequence:
\begin{equation}
\textstyle    \chi(T) = \frac{1}{\pi} \int_{0}^{\infty} \frac{S(\omega)}{\omega^2} F(\omega, T) \, d\omega
    \label{eq:chi_integral}
\end{equation}
This integral involves two key components: the \textbf{noise spectral density (NSD)}, $S(\omega)$, and the DD sequence's \textbf{filter function}, $F(\omega, T)$. $S(\omega)$ describes the distribution of the environmental noise across different frequencies $\omega$. Following common models for NV center nuclear spin bath~\cite{maze2013decoherence}, we model the NSD with a Gaussian profile superimposed on a constant offset~\cite{hernandez18,Martina_2023}:
\begin{equation}
\textstyle    S(\omega) = y_0 + a \exp\left(-\frac{(\omega-v_L)^{2}}{2w_1^{2}}\right)
    \label{eq:gaussian_nsd}
\end{equation}
Here, $y_0$ is a noise floor, and the Gaussian term captures the primary noise peak characterized by amplitude $a$, center frequency $v_L$, and width $w_1$. The filter function, $F(\omega, T)$, characterizes how effectively the applied DD sequence suppresses noise at frequency $\omega$. It is determined by the sequence's pulse timings. These timings are encoded in the \textbf{modulation function}, $y(t)$, which represents the effective time-dependent switching (+1 or -1) of the qubit-noise interaction induced by the sequence's $\pi$ pulses. The filter function $F(\omega, T)$ is defined as the squared magnitude of the Fourier transform, $Y(\omega, T)$, of this modulation function:
\begin{equation}
    F(\omega, T) = |Y(\omega, T)|^2,
    \label{eq:filter_function}
\end{equation}
where the Fourier transform is given by:
\begin{equation}
    Y(\omega, T) = \textstyle\int_{0}^{T} y(t) e^{-i\omega t} \, dt.
    \label{eq:fourier_transform}
\end{equation}

Standard forms for the modulation function $y(t)$, corresponding to sequences like FID (no pulses, $y(t)=1$), Hahn echo (one pulse), CPMG (evenly spaced pulses), and UDD (non-uniformly spaced pulses), provide the building blocks for constructing and analyzing the sequences generated in this work (see Fig.~\ref{fig:pulse_shapes})~\cite{hahn1950spin,meiboom1958modified,uhrig2007keeping}. By calculating $y(t)$, $Y(\omega, T)$, $F(\omega, T)$, and using the model for $S(\omega)$, we can compute $\chi(T)$ via Eq.~\ref{eq:chi_integral} and subsequently the coherence $W(T)$. Table~\ref{tab:coherence_params} summarizes the parameters used in this formalism.

Designing an optimal DD pulse sequence for a given NV center and noise environment is non-trivial. Different sequence patterns implement different filter functions, and the optimal pattern depends on the noise spectral density of the NV’s environment. In diamond, the spin-bath-induced noise often has a broadband or Gaussian spectral profile (due to many thermally driven $^{13}$C spins)~\cite{doherty2013nitrogen}, and no single analytical sequence (CPMG, UDD, etc.) is guaranteed to be optimal in all conditions. Moreover, practical constraints such as a fixed total sequence duration or limited timing resolution further restrict the space of viable sequences.

\begin{table}[t]
\centering
\caption{Summary of coherence computation parameters.}
\label{tab:coherence_params}
\vspace{-3mm}
\scalebox{0.88}{
\begin{tabular}{>{\columncolor{blue!20}}l>{\columncolor{blue!5}}l}
\toprule
\textbf{Symbol} & \textbf{Description} \\
\midrule
$T$ & Total evolution time \\
$W(T)$ & Final qubit coherence at time $T$ \\
$\chi(T)$ & Decoherence function \\
$\omega$ & Angular frequency component of noise \\
$S(\omega)$ & Noise spectral density (NSD) \\
$F(\omega, T)$ & Filter function of DD sequence \\
$y(t)$ & Modulation function encoding pulse sequence \\
$Y(\omega, T)$ & Fourier transform of $y(t)$ \\
$y_0$ & Constant noise floor \\
$a$ & Amplitude of Gaussian noise peak \\
$v_L$ & Center frequency of Gaussian noise peak \\
$w_1$ & Width (std. dev.) of Gaussian noise peak \\
\bottomrule
\end{tabular}}
\vspace{-3mm}
\end{table}

\begin{table}[t]
\centering
\caption{Summary of Q-learning parameters used in Sec.~\ref{sec:qlearn}.}
\label{tab:rl_params}
\vspace{-3mm}
\scalebox{0.88}{
\begin{tabular}{>{\columncolor{purple!10}}l>{\columncolor{pink!10}}l>{\columncolor{purple!10}}l>{\columncolor{pink!10}}l}
\toprule
\textbf{Symbol} & \textbf{Description} & \textbf{Symbol} & \textbf{Description} \\
\midrule
$s$ & Current state & $s'$ & Next state \\
$S$ & Set of all states & $a$ & Action taken in $s$ \\
$a'$ & Candidate next action & $A$ & Set of all actions \\
$r$ & Reward for $(s, a)$ & $r_{t+1}$ & Reward at time $t+1$ \\
$R$ & Reward distribution & $t$ & Discrete time index \\
$Q(s, a)$ & Action-value function & $Q^*(s, a)$ & Optimal action-value \\
$\alpha$ & Learning rate & $\gamma$ & Discount factor \\
$\mathbb{E}$ & Expectation operator &  &  \\
\bottomrule
\end{tabular}}
\vspace{-5mm}
\end{table}


\subsection{Reinforcement Learning and Q-Learning}
\label{sec:qlearn}

Reinforcement Learning (RL) provides a natural framework for solving sequential decision-making problems such as constructing dynamical decoupling (DD) sequences. In RL, an agent interacts with an environment over time by observing a \textbf{state}, taking an \textbf{action}, and receiving a scalar \textbf{reward}. The agent learns a \textbf{policy} that maps states to actions to maximize the cumulative reward. Unlike supervised learning, RL requires no labeled data; learning proceeds via trial and error based solely on observed rewards.

Q-learning~\cite{watkins1992q}, a widely used model-free RL algorithm, learns the optimal action-value function $Q^*(s, a)$ (refer to Table~\ref{tab:rl_params} for parameter references): the expected total return when taking action $a$ in state $s$ and following the optimal policy thereafter. For problems with discrete states and actions, this function maps state-action pairs to their estimated values. Q-learning updates its estimates from transitions $(s_t, a_t, r_{t+1}, s_{t+1})$ by:
\begin{equation}
\textstyle    Q(s_t, a_t) \leftarrow Q(s_t, a_t) + \alpha \left[ r_{t+1} + \gamma \max_{a' \in A} Q(s_{t+1}, a') - Q(s_t, a_t) \right],
    \label{eq:q_update}
\end{equation}
where $\alpha$ is the learning rate, and $\gamma$ is the discount factor for future rewards. The agent refines $Q(s, a)$ estimates over time by minimizing the temporal difference error between the current estimate and the target value. Q-learning attempts to iteratively achieve the optimal Q-function, which satisfies the Bellman equation:
\begin{equation}
\textstyle    Q^*(s, a) = \mathbb{E}\left[ r + \gamma \max_{a' \in A} Q^*(s', a') \mid s, a \right],
\end{equation}
where the expectation is over the reward distribution and stochastic transition dynamics. A central challenge in Q-learning is balancing \textbf{exploration} (trying unseen actions) with \textbf{exploitation} (choosing known high-value actions). \textbf{Epsilon-greedy} policies are commonly used to manage this trade-off. Due to its generality, ability to operate in unknown environments, and support for discrete action sequences, Q-learning is a compelling choice for optimizing DD segment selection to preserve quantum coherence.

%% file: sections/motivation.tex
\section{Motivation for \sol{}}
\label{sec:motivation}

NV centers' sensing potential is hindered by decoherence; maximizing the coherence time is, therefore, a critical objective, as longer coherence directly translates to enhanced sensitivity in sensing applications. Dynamical decoupling sequences, employing precisely timed microwave pulses, represent one of the main techniques for combating this decoherence. Standard sequences like Hahn echo, CPMG, and UDD have proven highly effective, extending coherence times by filtering out environmental noise. Despite their success, however, these sequences face significant limitations. They are typically designed based on assumptions about the noise (e.g., specific characteristics) and are, therefore, not universally optimal. The actual noise environment experienced by an individual NV center can deviate significantly from idealized models \cite{bauch2020,romach2015}. Consequently, applying a standard DD sequence often results in suboptimal coherence protection for the specific noise profile encountered \cite{Alvarez2011}.

\begin{figure}[t]
    \centering
    \includegraphics[width=0.99\linewidth]{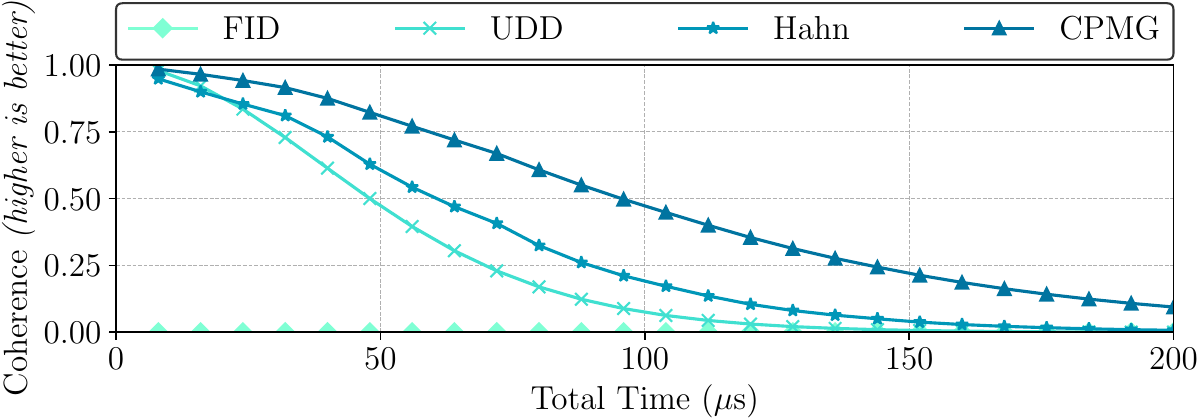}
    \vspace{-3mm}
    \caption{Standard DD sequences have rapidly decaying coherence over time -- even the best performing ones like CPMG.}
    \vspace{-5mm}
    \label{fig:motivation}
\end{figure}

This can be observed in Fig.~\ref{fig:motivation}, which shows the mean coherence of each DD sequence (when only the respective DD sequence is applied for the entire evolution duration) across 1000 different noise scenarios -- refer to Sec.~\ref{sec:methodology} for more details on the methodology. As the figure shows, the sequences decohere rapidly, approaching the near-zero coherence of applying no DD, as reflected in the Free Induction Decay (FID) line. Note: the FID curve appears to start at 0 because it rapidly decoheres due to the lack of DD techniques. Thus, the limitations of analytically derived DD sequences highlight a major challenge: designing pulse sequences that are optimal for specific, complex, and unknown noise environments. Furthermore, the number of distinct dynamical decoupling pulse sequences grows exponentially with the total evolution time, making it infeasible to consider all possibilities with brute-force methods or conventional optimization techniques. 

\textit{Thus, there is a need for novel methods capable of efficiently discovering DD sequences tailored to complex noise environments. This work addresses this challenge by leveraging RL, as we describe next.}

%% file: sections/design.tex
\section{\sol{}'s Design and Implementation}
\label{sec:design}

\begin{figure*}[t]
    \centering
    \includegraphics[width=0.99\textwidth]{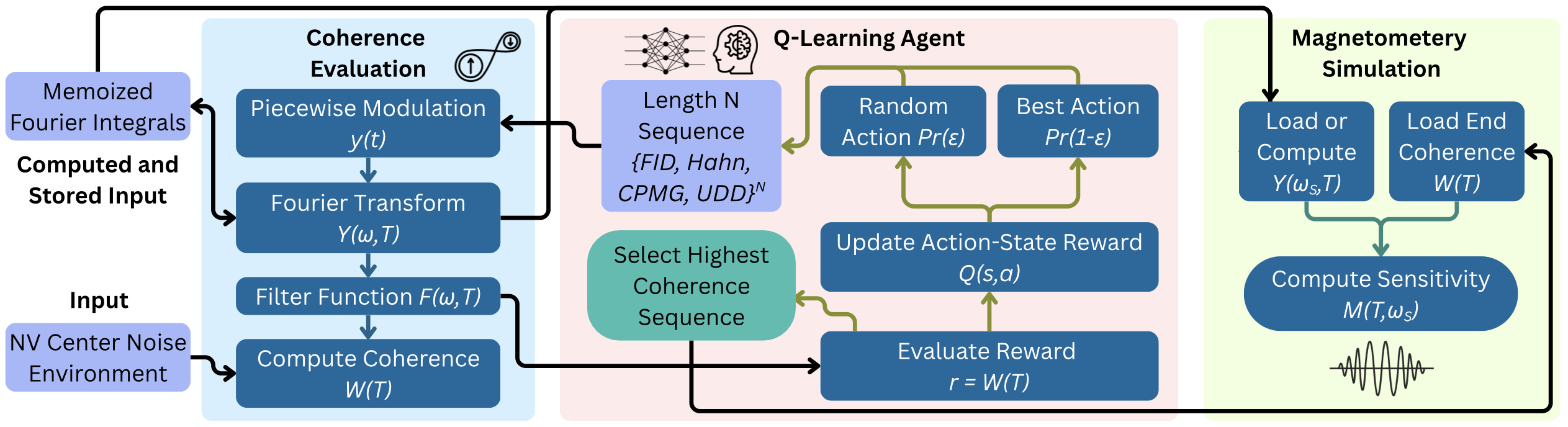}
    \vspace{-3mm}
    \caption{Overview of the \sol{} framework. Our system has three main components: the Coherence Evaluation pipeline, which computes the coherence using filter functions; the Q-Learning agent, which uses reinforcement learning to select optimal piecewise DD sequences; and the Magnetometry Simulation, which evaluates our AC magnetometry sensitivity metric.}
    \vspace{-5mm}
    \label{fig:overview}
\end{figure*}

\textbf{System Workflow.} \sol{}'s workflow is an end-to-end RL framework for discovering high-fidelity DD sequences that extend NV center coherence and enhance quantum sensing performance. It makes three key contributions (Fig.~\ref{fig:overview}). First, it formulates DD synthesis as a sequential decision-making task, enabling a Q-learning agent to learn optimal piecewise sequences without requiring explicit noise modeling. Second, it introduces a modular coherence evaluation pipeline that uses segment-wise Fourier transforms with memoization to efficiently simulate large numbers of DD candidates. Third, it derives a sensitivity-aware metric that links coherence and frequency-domain filtering to NV magnetometry performance. Together, these elements enable \sol{} to learn adaptive DD protocols offline and generalize across noise settings.


\subsection{\sol{}'s Q-learning Agent}

The core of \sol{} is a reinforcement learning agent designed to discover optimal DD sequences that maximize NV center coherence over a specified total evolution time $T$ (refer to Sec.~\ref{sec:background} and Table~\ref{tab:rl_params} for a background overview on RL). To make the learning process computationally tractable, the total time $T$ is discretized into $N = \frac{T}{4\,\mu\text{s}}$ segments of fixed duration (4~\textmu s). In each segment $k$ (from $1$ to $N$), the agent must choose an \textbf{action}, $a_k$, corresponding to one of four predefined DD subsequence types: Free Induction Decay (FID), Hahn echo (Hahn), Carr-Purcell-Meiboom-Gill (CPMG), or Uhrig Dynamical Decoupling (UDD). These actions are encoded numerically, for instance, using the set $a_k \in \{0, 1, 2, 3\}$.

One aspect of the RL formulation is the definition of the \textbf{state}, $s$, which represents the information the agent uses to select the next action. A naive approach would define the state after $k$ segments as the complete history of actions taken, $s_k = (a_1, a_2, \ldots, a_k)$. However, the number of such states grows exponentially with the sequence length, quickly leading to an unmanageably large state space for practical values of $N$. To overcome this, we employ \textbf{state aggregation} using a truncated history. The state relevant for choosing the action $a_{k+1}$ is defined by only the $m$ most recent actions: $s'_k = (a_{k-m+1}, \ldots, a_{k-1}, a_k)$, where $m$ is a small, fixed integer determining the memory length ($m=3$ in our experimentation). If the sequence length $k$ is less than $m$, the state tuple $s'_k$ is padded with a special value (e.g., $-1$) at the beginning to ensure a consistent state representation dimension. For instance, with $m=2$, the sequence of states used for decision-making evolves as $s'_0 = (-1, -1)$, $s'_1 = (-1, a_1)$, $s'_2 = (a_1, a_2)$, $s'_3 = (a_2, a_3)$, and so on. This approach bounds the total number of states to $O(4^m)$, making the learning problem scalable regardless of the total sequence length $N$. It relies on the intuition that the optimal choice for the next segment depends more on the recent pulse history than on the earlier one.

The agent interacts with the environment over an \textbf{episode}, which consists of sequentially choosing $N$ actions to construct a complete DD sequence $Seq_N = (a_1, a_2, \ldots, a_N)$. At each step $k$ (from $0$ to $N-1$), the agent observes the current aggregated state $s'_k$ and selects the next action $a_{k+1}$ according to an \textbf{epsilon-greedy policy}. This policy balances exploration and exploitation:
\[
a_{k+1} =
\begin{cases}
\text{a random choice from } \{0,1,2,3\}, & \text{with probability } \epsilon, \\
\arg\max_{a' \in \{0,1,2,3\}} Q(s'_k, a'), & \text{with probability } 1 - \epsilon.
\end{cases}
\]
Here, $Q(s', a')$ is the agent's estimate of the expected future reward for taking action $a'$ in state $s'$, and $\epsilon$ is the exploration rate, which is typically annealed (gradually decreased) over the course of training from a high value (e.g., 1) to a low value (e.g., 0.03).

Once the full sequence $Seq_N = (a_1, \ldots, a_N)$ is completed, the candidate sequence is then evaluated to calculate the final NV spin coherence, $W$. The coherence of the sensor with the candidate sequence serves as the \textbf{reward}, $r$, for the entire episode:
\[
r = W = \exp(-\chi),
\]
where $\chi$ is the decoherence function computed from the overlap between the sequence's filter function (derived from its modulation function) and the noise spectral density of the environment (refer to Sec.~\ref{sec:background} and Table~\ref{tab:coherence_params} for background on computing coherence).

We use a Monte Carlo approach to update the action-value estimates. After each episode concludes and the final reward $r$ is obtained, the Q-value for every state-action pair $(s'_k, a_{k+1})$ encountered during that episode is updated towards the observed outcome:
\[
Q(s'_k, a_{k+1}) \leftarrow Q(s'_k, a_{k+1}) + \alpha \Bigl( r - Q(s'_k, a_{k+1}) \Bigr),
\]
where $\alpha$ is the learning rate (a small positive constant, e.g., 0.1). This update rule adjusts the agent's expectation for taking action $a_{k+1}$ in the aggregated state $s'_k$ based on the actual final coherence achieved in that episode. Over many episodes, the Q-values converge, allowing the agent to learn an effective policy for constructing high-coherence sequences.

After sufficient training, the agent's learned policy can be used to construct the best-found DD sequence for a given noise setting. This is done by starting from the initial state $s'_0$ and greedily selecting the action with the highest Q-value at each step, without exploration ($\epsilon=0$):
Let $\hat{Seq} = (a_1^*, \ldots, a_N^*)$ be the resulting sequence actions. For $k=0, \ldots, N-1$: $a_{k+1}^* = \text{argmax}_{a' \in \{0,1,2,3\}} Q\bigl( s'_k, a' \bigr)$, where $s'_k$ is the aggregated state derived from the partial sequence $(a_1^*, \ldots, a_k^*)$ constructed so far. This final sequence $\hat{Seq}$ represents the optimized DD protocol discovered by the RL agent.

\begin{figure}[t]
    \centering
    \includegraphics[width=0.99\linewidth]{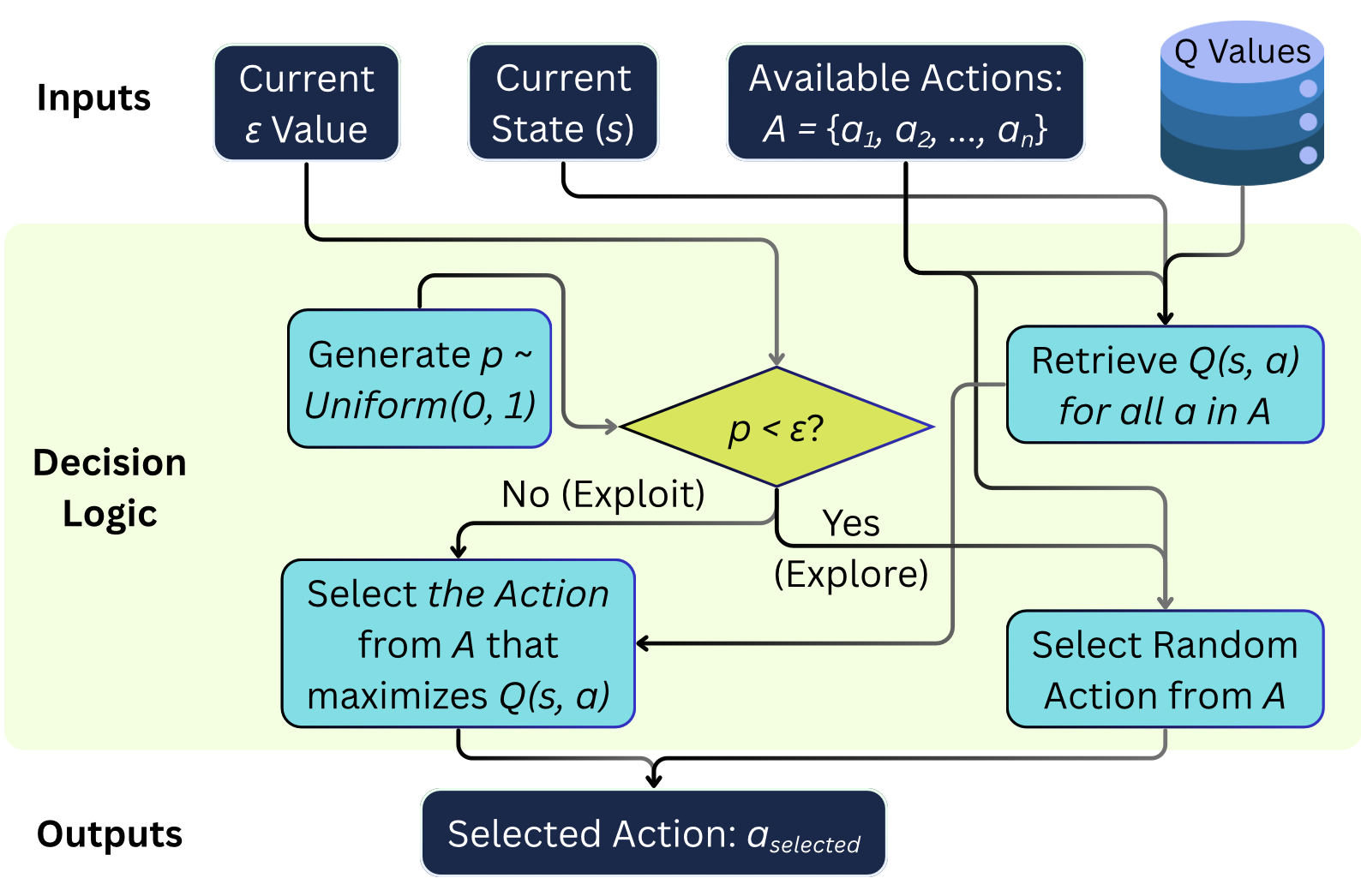}
    \vspace{-3mm}
    \caption{The decision logic for \sol{}'s Q-Learning agent.}
    \vspace{-5mm}
    \label{fig:main_dag}
\end{figure}

Finally, to accelerate learning when optimizing sequences for progressively longer total times $T$, we leverage the concept of \textbf{warm-starting}. If the agent has already found an optimal sequence $(a_1^*, \ldots, a_k^*)$ for a shorter duration corresponding to $k$ segments, this sequence is used as the \textit{base sequence} when learning the sequence for $N > k$ segments. The agent begins its episode construction with this base sequence, determines the initial aggregated state $s'_k$ from its end, and only learns the optimal actions $a_{k+1}, \ldots, a_N$ for the remaining $N-k$ segments. This focuses the learning effort on extending the known good solution, leveraging prior knowledge, and significantly reducing the effective search space. Refer to Fig.~\ref{fig:main_dag} for the execution flow of this process.

The Q-learning framework designed for \sol{}, enhanced with state aggregation and warm-starting, provides a scalable method for discovering complex, high-performance DD sequences tailored to specific noise environments, without requiring an explicit analytical noise model. The steps for the Q-learning algorithm are summarized in Algorithm~\ref{alg:qlearning}. The agent learns directly from the outcomes of its actions, guided by the reward signal derived from the final NV center coherence. The success and efficiency of this learning process, therefore, critically depend on the rapid evaluation of the coherence $W(T)$ for each candidate sequence generated during training. The pipeline developed within \sol{} to perform this complex computation is detailed in the following section.

\subsection{\sol{}'s Coherence Evaluation Pipeline}

The Q-learning agent described above requires frequent evaluation of candidate DD sequences to obtain the coherence reward $W(T)$ that drives its learning process. Directly calculating $W(T)$ using the filter function formalism described in Sec.~\ref{sec:background} presents a significant computational bottleneck for RL training. The core challenge lies in determining the sequence's filter function $F(\omega, T) = |Y(\omega, T)|^2$; this necessitates computing the Fourier transform $Y(\omega, T)$ via numerical integration (Eq.~\ref{eq:fourier_transform}) repeatedly across a wide range of frequencies $\omega$ to accurately evaluate the decoherence integral (Eq.~\ref{eq:chi_integral}). Performing these intensive calculations for every sequence explored during thousands of RL training episodes would be prohibitively expensive and computationally intensive. To make the RL approach feasible, \sol{} incorporates a highly efficient coherence evaluation pipeline built upon two key implementation strategies: piecewise Fourier transforms and memoization.

\vspace{2mm}

\noindent\textbf{Piecewise Fourier Transform Calculation.} As \sol{} constructs DD protocols ($Seq$) piecewise from a sequence of $N$ basic segments ($a_k \in \{\text{FID, Hahn, CPMG, UDD}\}$), each of duration $\Delta t$, the overall modulation function $y(t)$ is also piecewise (for our experimentation, $\Delta t=4 \mu s$). Leveraging the linearity of the Fourier transform (Eq.~\ref{eq:fourier_transform}), the total transform $Y(\omega, T)$ needed for the filter function (Eq.~\ref{eq:filter_function}) is computed by summing the individual Fourier transforms $Y_k(\omega)$ of each segment $k$ as follows:
\begin{equation}
Y(\omega, T) = \textstyle\sum_{k=1}^{N} Y_k(\omega) = \textstyle\sum_{k=1}^{N} \textstyle\int_{(k-1)\Delta t}^{k\Delta t} y_k(t') e^{-i\omega t'} \, dt'.
\end{equation}
Here, $y_k(t')$ is the modulation function specific to the type and internal parameters of the $k$-th segment. Each segment's transform $Y_k(\omega)$ is calculated via numerical integration over its respective time interval. This piecewise approach breaks down the computation into manageable, modular units corresponding to actions of \sol{} Q-learning agents.

\vspace{2mm}

\begin{algorithm}[t]
\caption{\sol{}'s Q-learning algorithm.}
\label{alg:qlearning}
\begin{algorithmic}[1]
    \State \textbf{Inputs:} $T, \Delta t, m, \mathcal{A}=\{0..3\}, N_{eps}, \alpha, \epsilon_{params}, Seq_{base}$
    \State \textbf{Initialize:} $N \!=\! T / \Delta t$; $k_{st} \!=\! \text{len}(Seq_{base})$; $N_{xtra} \!=\! N \!-\! k_{st}$
    \State \quad $Q(s', a') \gets 0 \; \forall s', a'$; $\epsilon \gets \epsilon_{start}$
    \For{episode $\gets 1$ to $N_{eps}$}
        \State $Seq \gets \text{list}(Seq_{base})$; $hist \gets []$
        \For{$k \gets k_{st}$ to $N-1$}
            \State $s'_k \gets \text{AggregateState}(Seq, m)$
            \If{rand() $< \epsilon$} $a_{k+1} \!\gets\! \text{rand}(\mathcal{A})$
            \Else ~$a_{k+1} \!\gets\! \arg\max_{a'} Q(s'_k, a')$ \EndIf
            \State Append $(s'_k, a_{k+1})$ to $hist$
            \State Append $a_{k+1}$ to $Seq$
        \EndFor
        \State $r \gets \text{Evaluate}(Seq)$
        \For{$(s', a)$ in $hist$}
            \State $Q(s', a) \gets Q(s', a) + \alpha ( r - Q(s', a) )$
        \EndFor
        \State $\epsilon \gets \max(\epsilon_{end}, \epsilon \times \epsilon_{decay})$
    \EndFor
\end{algorithmic}
\end{algorithm}

\noindent\textbf{Utilizing Memoization for Efficient Computation.} The primary computational cost lies in the repeated numerical integration required to find $Y_k(\omega)$ for each segment $k$ at every frequency $\omega$ of the coherence integral. Crucially, the value of $Y_k(\omega)$ for a given segment type, duration, absolute time interval, and frequency $\omega$ is deterministic and independent of the noise spectral density $S(\omega)$.

\sol{} exploits this by employing \textbf{memoization}. The numerical result for $Y_k(\omega)$ is computed only once for each unique combination of segment parameters and frequency $\omega$. This result is stored in a persistent cache (e.g., a nested dictionary). Subsequent requests to calculate $Y_k(\omega)$ for the exact same configuration retrieve the stored complex value directly, completely bypassing the expensive numerical integration. Since the RL agent frequently explores sequences containing identical segments at the same positions (especially with state aggregation), this strategy drastically reduces redundant computations, particularly when evaluating many sequences or the same sequence against different noise profiles. This memoization is the key optimization that makes the extensive simulations required for RL training computationally tractable.

This pipeline, combining the piecewise calculation of $Y(\omega, T)$ with memoized segment transforms $Y_k(\omega)$, enables rapid computation of the filter function $F(\omega, T)$. This filter function is then numerically integrated against the specified noise spectral density $S(\omega)$ to find the decoherence function $\chi(T)$. The final coherence $W(T) = \exp(-\chi(T))$ is then returned to the Q-learning agent as the reward signal $r$, enabling efficient learning cycles. Thus, the coherence evaluation pipeline enables \sol{} to discover sequences yielding high coherence $W(T)$. We now analyze how the learned DD sequences translate into improved sensitivity in AC magnetometry with NV centers.

\subsection{\sol{}'s Magnetometry Simulation}

While \sol{} effectively optimizes DD sequences for maximal coherence $W(T)$, extending its capability to enhance specific quantum sensing tasks, such as NV center AC magnetometry, requires a more targeted approach. NV center AC magnetometry enables the detection of extremely weak, time-varying magnetic fields with high resolution, with applications such as imaging biological processes at the sub-cellular level and characterizing magnetic materials~\cite{Casola2018,BALASUBRAMANIAN201469}. Standard DD optimization often overlooks that peak sensitivity for NV centers ($\eta$) depends not only on coherence but also on the sequence's frequency response at the target signal frequency $\omega_s$.

To bridge this gap and enable \sol{} to directly optimize NV center sensing performance, we derive a practical, sensitivity-aware metric. This metric explicitly connects the achievable AC magnetometry sensitivity $\eta$ to the fundamental properties of the DD sequence evaluated by our pipeline: the final NV center coherence $W(T)$ and the magnitude of the modulation function's Fourier transform, $|Y(\omega_s, T)|$, at the frequency $\omega_s$.


First, we derive the scaling relationship for the shot-noise limited sensitivity $\eta$ (units: Field/$\sqrt{\text{Hz}}$) of AC magnetometry using an NV center sensor undergoing a DD sequence of duration $T$.

The standard sensitivity $\eta$ relates the sensitivity achieved in a single measurement cycle of duration $T$ ($\eta_{\text{shot}}$, units: Field) to the sensitivity per unit bandwidth by accounting for averaging over a total time $T_{\text{total}} = M \times T$:
\begin{equation}
    \eta = \eta_{\text{shot}} \sqrt{T}
    \label{eq:eta_def}
\end{equation}
This scaling arises because averaging $M$ independent measurements reduces the noise standard deviation by $\sqrt{M}$, and sensitivity per $\sqrt{\text{Hz}}$ normalizes for the total measurement time $T_{\text{total}}$\cite{taylor2008}.

An NV center's magnetometer signal is dependent linearly on the strength of b, the magnetic field being measured $\mathcal{S} \approx \frac{g\mu_B}{\hbar}\,b\,\tau$.
Since S and b have a linear relationship: $\frac{d\mathcal{S}}{db} = \frac{g\mu_B}{\hbar}\,\tau$ and $\left|\frac{d\mathcal{S}}{db}\right|_{\text{max}} = \frac{g\mu_B}{\hbar}\,\tau$.
For a sufficiently small change in the magnetic field $\delta b$ to be captured by our NV center's signal $S$, it must be at least comparable to the noise in the signal, $\sigma_S$, meaning that the minimum detectable change in signal is $\delta S \approx \sigma_S$. By equating a change in the signal $\delta S = \left|\frac{d\mathcal{S}}{db}\right|_{\text{max}} \delta b $to the noise $\sigma_S$, we get $\sigma_S \approx \left|\frac{d\mathcal{S}}{db}\right|_{\text{max}}\,\delta b\quad\Longrightarrow\quad \delta b = \frac{\sigma_S}{\left|\frac{d\mathcal{S}}{db}\right|_{\text{max}}}$. Defining the single-shot sensitivity as the minimum detectable magnetic field change, we obtain:
\begin{equation}
    \eta_{\text{shot}} = \frac{\sigma_S}{\left|\frac{d\mathcal{S}}{db}\right|_{\text{max}}}\
    \label{eq:delsig}
\end{equation}

In a typical NV-based AC magnetometry setup, the measured signal $S$ depends on the phase $\phi$ accumulated by the NV spin due to the external field, $b$. This signal is inherently limited by the remaining coherence $W(T)$ after the sensing interval $T$, such that $S \approx W(T)\cos(\phi + \phi_{\text{bias}})$. Note that the accumulated phase $\phi$ is not only influenced by the external field but also by how the DD sequence modulates the NV center's interaction with this field over time, described by the modulation function $y(t)$.


Next, in a typical Ramsey measurement, a qubit is first prepared in an equal superposition state, $\ket{\psi_0} = \frac{1}{\sqrt{2}}\Bigl(\ket{0} + \ket{1}\Bigr)$, and then allowed to evolve freely for a time \(T\) during which it accumulates a relative phase \(\phi\) due to an external field. In the ideal, decoherence-free case, the state after evolution becomes $\ket{\psi(T)} = \frac{1}{\sqrt{2}}\Bigl(\ket{0} + e^{-i\phi}\ket{1}\Bigr)$.  A final \(\pi/2\)-pulse is then applied, typically with an additional phase 
offset \(\phi_{\text{bias}}\), which rotates the state into a basis in which the observable is measured. The expectation value of the measurement operator is given by~\cite{degen2017}: $\langle\sigma_z\rangle = \cos\Bigl(\phi + \phi_{\text{bias}}\Bigr)$. Of course, in practice, the aforedescribed effects of decoherence apply to the qubit during this evolution, such that the measured signal becomes: $\langle \sigma_z\rangle = S = W(T)\cos\Bigl(\phi + \phi_{\text{bias}}\Bigr)$, where $W(T) = \exp[-\chi(T)]$ (the coherence), as mentioned above.

We consider an AC magnetic field signal, which can be generally described by the sinusoidal form $B(t) = b \cos(\omega_s t + \phi_s)$, with \(b=1\) (i.e., the phase shift is given per unit field amplitude). This magnetic field induces a phase $\phi = \frac{\gamma}{2} \int_{0}^{T} y(t')\,B(t')\,dt'$, where \(y(t)\) is the DD modulation function and \(\gamma\) is the (constant) gyromagnetic ratio~\cite{taylor2008,degen2017}.  Substituting \(B(t')\) we obtain $\phi = \frac{\gamma}{2} \int_{0}^{T} y(t')\,\cos(\omega_s t' + \phi_s)\,dt'$.

Using Euler's formula, we express the cosine as the real part of an exponential: $\cos(\omega_s t' + \phi_s) = \Re\Bigl\{e^{i(\omega_s t' + \phi_s)}\Bigr\}$. Thus, $\phi = \frac{\gamma}{2}\,\Re\left\{ e^{i\phi_s}\int_{0}^{T} y(t')\,e^{i\omega_s t'}\,dt'\right\}$. By definition, the Fourier transform of \(y(t)\) at frequency \(\omega_s\) is $Y(\omega_s,T) = \int_{0}^{T} y(t')\,e^{-i\omega_s t'}\,dt'$. Since \(y(t')\) is real, we have $\int_{0}^{T} y(t')\,e^{i\omega_s t'}\,dt' = Y^*(\omega_s,T)$, where $Y^*$ is the complex conjugate of $Y(\omega_s,T)$. Hence, the phase can be written as $\phi = \frac{\gamma}{2}\,\Re\Bigl\{e^{i\phi_s}\,Y^*(\omega_s,T)\Bigr\}$. It is important to step back and recognize that the goal is to maximize the sensitivity of the accumulated phase in our sensor to the magnetic field strength, i.e., $\left|\frac{d\phi}{db}\right|$. To maximize phase accumulation, we choose the overall reference phase so that $e^{i\phi_s}=1$. Then the phase amplitude becomes $\phi_{\text{amp}} \approx \frac{\gamma}{2}\,\Re\Bigl\{Y^*(\omega_s,T)\Bigr\} 
= \frac{\gamma}{2}\,\Bigl|Y(\omega_s,T)\Bigr|$, which is equivalent to $\left|\frac{d\phi}{db}\right|_{\text{max}} \approx \frac{\gamma}{2}\,\Bigl|Y(\omega_s,T)\Bigr|$, since we have set \(b=1\).

Since $S$ depends on phase $\phi$, and $\phi$ on magnetic field amplitude $b$ (e.g. $S = S(\phi)$, $\phi = \phi(b)$), we apply the chain rule: $\frac{dS}{db} = (\frac{dS}{d\phi}) (\frac{d\phi}{db})$. Evaluating the terms of the chain rule, first, we differentiate $S$ w.r.t $\phi$. Using the signal model $S = W(T)\cos(\phi + \phi_{\text{bias}})$ (ignoring any constant offset $S_0$ which vanishes upon differentiation):
\begin{equation}
\textstyle \frac{dS}{d\phi} = \frac{d}{d\phi} \left[ W(T)\cos(\phi + \phi_{\text{bias}}) \right] = -W(T)\sin(\phi + \phi_{\text{bias}})\notag
\end{equation}
Substituting this and the expression for $\frac{d\phi}{db}$ into the chain rule gives: $\frac{dS}{db} = \left[ -W(T)\sin(\phi + \phi_{\text{bias}}) \right] \left[ \frac{d\phi}{db} \right]$. We are interested in the maximum slope magnitude, which determines the sensitivity. The magnitude is: $\left|\frac{dS}{db}\right| = W(T) \left|\sin(\phi + \phi_{\text{bias}})\right| \left|\frac{d\phi}{db}\right|$.

To maximize this, we choose the operating point (by adjusting $\phi_{\text{bias}}$ or measuring at the point of maximum slope) such that $|\sin(\phi + \phi_{\text{bias}})| = 1$. We also use the maximum value for the phase sensitivity derived earlier, $\left|\frac{d\phi}{db}\right|_{\text{max}} \approx \frac{\gamma}{2} |Y(\omega_s, T)|$. This yields the maximum signal slope: $\left|\frac{dS}{db}\right|_{\text{max}} = W(T) \times 1 \times \left|\frac{d\phi}{db}\right|_{\text{max}} \approx W(T) \frac{\gamma}{2} |Y(\omega_s, T)|$

Substitute the maximum slope into the single-shot sensitivity definition (Eq.~\ref{eq:delsig}):
\begin{equation}
    \textstyle \eta_{\text{shot}} \approx \frac{\sigma_S}{\frac{\gamma}{2} W(T) |Y(\omega_s, T)|}\notag
    \label{eq:eta_shot_detail}
\end{equation}
Now, use the relationship $\eta = \eta_{\text{shot}} \sqrt{T}$ (Eq.~\ref{eq:eta_def}) to find the sensitivity per unit bandwidth:
\begin{equation}
    \textstyle \eta = \eta_{\text{shot}} \sqrt{T} \approx \left( \frac{\sigma_S}{\frac{\gamma}{2} W(T) |Y(\omega_s, T)|} \right) \sqrt{T}\notag
\end{equation}
We assume that measurement noise $\sigma_S$ is constant for our experimentation. Rearranging and grouping constant terms:
\begin{equation}
    \textstyle \eta \approx \underbrace{\left( \frac{2 \sigma_S}{\gamma} \right)}_{\text{Constants}} \times \frac{\sqrt{T}}{W(T) |Y(\omega_s, T)|}\notag
    \label{eq:eta_final_detail}
\end{equation}

Therefore, the scaling of sensitivity with respect to the sequence properties is:
\begin{equation}
    \textstyle \eta \propto \frac{\sqrt{T}}{W(T) |Y(\omega_s, T)|}\notag
    \label{eq:sensitivity_scaling_final}
\end{equation}
Based on this proportionality, we define the \textbf{Relative Sensitivity Metric} $M$ used in our simulations:
\begin{equation}
    \textstyle M(T, \omega_s) = \frac{\sqrt{T}}{W(T) |Y(\omega_s, T)|}
    \label{eq:metric_definition}
\end{equation}
A lower value of $M$ indicates better sensitivity. This metric $M$ is directly proportional to the standard sensitivity $\eta$ (Field/$\sqrt{\text{Hz}}$), and thus provides a way to compare the potential sensing performance of different DD sequences based on simulation results. 
The significance of this derived metric $M$ lies in providing a computationally straightforward way to assess the AC magnetometry sensitivity potential of any candidate DD sequence. It directly utilizes the coherence $W(T)$ and the sequence's Fourier transform magnitude $|Y(\omega_s, T)|$ at the target signal frequency $\omega_s$. These are quantities readily available from the coherence evaluation pipeline described previously, often benefiting from the same memoization techniques used to accelerate the calculation of $Y(\omega, T)$ across many frequencies. \textit{This allows \sol{} to efficiently evaluate and optimize DD sequences not only for high coherence but also specifically for improved sensing performance, offering a practical advantage over methods requiring more complex environmental or use-case-specific simulations.}

%% file: sections/methodology.tex
\section{Experimental Methods}
\label{sec:methodology}

This section details \sol{}'s simulation environment, coherence calculation methods, RL implementation, and benchmarks.

\vspace{2mm}

\noindent\textbf{Simulation Environment.} All simulations and analyses were performed using Python 3.10.12. Core numerical computations relied on NumPy (v1.26.1)~\cite{harris2020array} and correlation analysis utilized SciPy (v1.15.2)~\cite{jones2016scipy}. The simulations were executed on a local research cluster equipped with 32-core 2.0 GHz AMD EPYC 7551P processors and 32 GB RAM, running Ubuntu 22.04.2 LTS. Custom Python modules were developed to handle noise generation, modulation functions, piecewise Fourier calculations with memoization, RL agent, and overall simulation control.


For our evaluations, we generated 1000 distinct NSD environments by sampling the parameters $(y_0, a, v_L, w_1)$ for the Gaussian family of noise models uniformly from ranges based on prior experimental noise spectroscopy studies~\cite{Martina_2023}. Specifically, the ranges used were: $y_0 \in [0.002, 0.008]$, $a \in [0.3, 0.7]$, $v_L$ corresponding to an effective magnetic field $B \in [520, 538]$~G (converted using $\gamma \approx 1.0705 \times 10^{-3} \text{ MHz/G}$), and $w_1 \in [0.004, 0.009]$. The integral for $\chi(T)$ (Eq.~\ref{eq:chi_integral}), involving $S(\omega)$ and the filter function $|Y(\omega, T)|^2$, is computed numerically over a range of frequencies $\omega$.

All DD sequences are constructed from $N = T/\Delta t$ contiguous segments, each of fixed duration $\Delta t = 4\,\text{\textmu s}$. We select this duration to achieve a manageable search while maintaining high sensitivity. The total Fourier transform $Y(\omega, T)$ for a complete piecewise sequence is calculated by summing the numerically computed Fourier transforms $Y_k(\omega)$ of each individual segment $k$, evaluated over its absolute time interval $[t_{\text{start}}, t_{\text{end}}]$. The numerical integration for each $Y_k(\omega)$ uses Numpy's trapz function, which follows the trapezoidal rule with 2000 points per 4\,\text{\textmu s} segment. The final numerical integration over frequency $\omega$ to compute decoherence $\chi$ considers the range from 0.001 to 8.5~rad/s based on prior work~\cite{Martina_2023}.

For efficiency, we implemented memoization for the segment Fourier transforms $Y_k(\omega)$, as discussed in Sec.~\ref{sec:design}. The result for each unique combination of segment type, segment parameters (e.g., $N=1$), absolute start/end times ($t_{\text{start}}, t_{\text{end}}$), and frequency ($\omega$) is stored in a nested dictionary structure upon first calculation. Subsequent requests for the same configuration retrieve the stored complex value, avoiding redundant numerical integration. This data is loaded from and saved to in-memory storage, enabling reuse of computations across runs and sequence evaluations. In addition, a pickle file containing the memoized dictionary's results is saved upon program completion.

\vspace{2mm}

\noindent\textbf{Q-learning Agent.} For state representation, a truncated history of the $m=3$ most recent actions (subsequences) is used for state aggregation, limiting the state space size to $4^3=64$. Special padding values are used initially when the history is shorter than $m$. The available actions $\mathcal{A}$ correspond to the four base DD types encoded numerically. The reward $r$ for a completed episode is the final calculated coherence $W(T) = \exp(-\chi(T))$. The learning rate is set to $\alpha = 0.1$. Exploration utilizes an epsilon-greedy policy with $\epsilon$ initially set to 1.0, decaying multiplicatively by $0.99$ per episode to a minimum of $\epsilon_{end} = 0.05$. Training proceeds for $N_{eps} = 300$ episodes per target time $T$ and NSD environment. To accelerate learning for longer sequences, warm-starting is used: the best sequence found for time $T$ serves as the fixed prefix (base sequence) when training for the subsequent time step $T + \Delta t$. While this may not necessarily be optimal, it empirically performs near-optimally (Sec.~\ref{sec:evaluation}) when compared to the Oracle technique described next.

\vspace{2mm}

\noindent\textbf{Oracle Technique.} Along with the standard DD pulses, we also evaluate \sol{} against an Oracle technique that finds the optimal pulse sequence using brute-force search at each time step ($4^N$ total choices where $N$ is the number of time segments in the pulse). While not realistic due to its prohibitive overhead, this Oracle technique helps establish an upper bound for performance. We also note that this is not the theoretical ideal, but the best that can be achieved with brute-force search under noise.

%% file: sections/evaluation.tex
\section{Evaluation, Analysis, and Discussion}
\label{sec:evaluation}

\begin{figure}[t]
    \centering
    \includegraphics[width=0.99\linewidth]{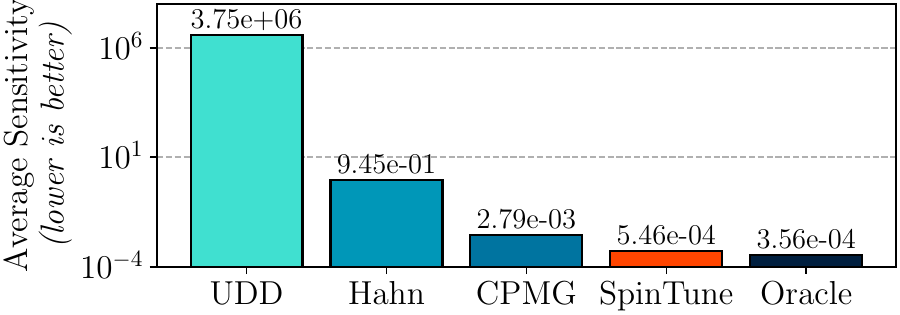}
    \vspace{-3mm}
    \caption{\sol{} has the lowest magnetometry sensitivity compared to standard DD sequences. Note the log scale.}
    \vspace{-4mm}
    \label{fig:sense_mag}
\end{figure}

\begin{figure}[t]
    \centering
    \includegraphics[width=0.99\linewidth]{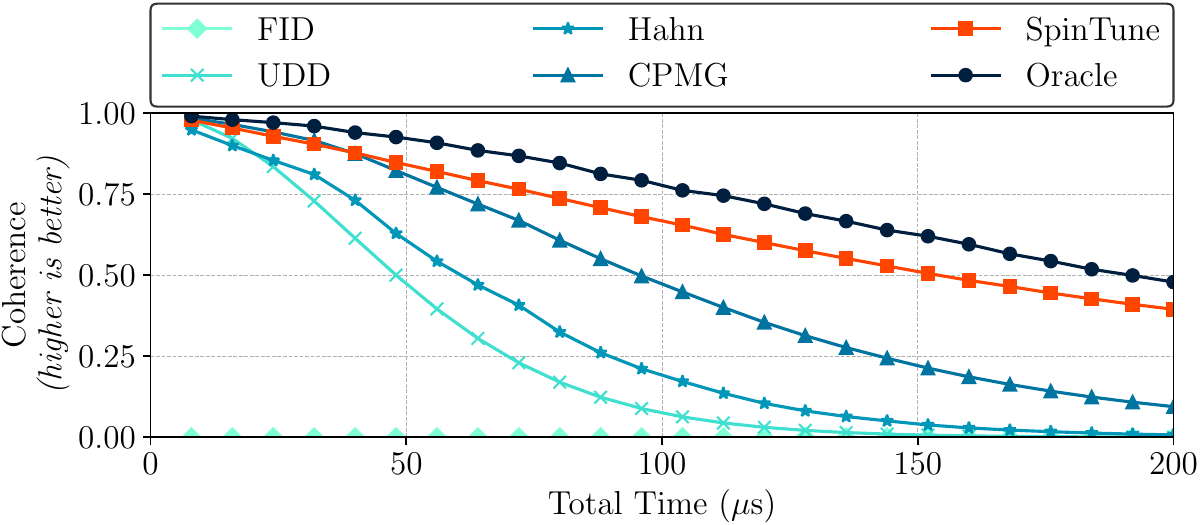}
    \vspace{-3mm}
    \caption{Compared to the use of baseline DD pulses, \sol{} consistently achieves much higher coherence that is close to the coherence of the Oracle technique.}
    \vspace{-5mm}
    \label{fig:main_perf}
\end{figure}

\noindent\textbf{(I) \sol{} demonstrates significantly improved magnetometry sensitivity compared to standard DD sequences, closely approaching the theoretical optimum.} We begin by presenting \sol{}'s flagship result and the end effect of its improvement in coherence. Fig.~\ref{fig:sense_mag} presents the calculated average AC magnetometry sensitivity, represented by the metric $M$ derived in Eq.~\ref{eq:metric_definition}, for the different DD strategies evaluated at an evolution time $T=200\,\text{\textmu s}$ and a target AC frequency of $\omega_s/2\pi = 1.0\,\text{MHz}$. Recall that $M \propto 1/(W(T)|Y(\omega_s, T)|)$, meaning a lower value indicates better magnetic field sensitivity. The results clearly demonstrate that the adaptive sequences generated by \sol{} yield significantly greater sensitivity than standard DD protocols. The performance hierarchy observed in the coherence plots is largely reflected in the sensitivity results, underscoring the role of coherence preservation in achieving high sensitivity. Compared to CPMG, the best-performing standard sequence, \sol{}, provides an approximate 5.1-fold improvement in sensitivity. Moreover, sequences such as Hahn and UDD exhibit relatively low sensitivity. The results highlight \sol{}'s ability to leverage its high coherence to achieve near-optimal AC magnetometry sensitivity. Next, we delve into how \sol{} achieves this.

\vspace{2mm}

\noindent\textbf{(II) On average, \sol{}'s learned dynamical decoupling sequence yields a 35\% point higher coherence when compared to the next best sequence tested, CPMG.} Fig.~\ref{fig:main_perf} presents the primary performance comparison, illustrating the mean coherence averaged over all simulated NSD environments as a function of total evolution time $T$ for the different dynamical decoupling strategies evaluated. As expected, coherence generally decays over time for all sequences employing pulses. The plot clearly demonstrates a performance hierarchy among the strategies. The standard Free Induction Decay (FID) sequence shows almost instantaneous coherence loss, which quickly approaches zero. The UDD sequence offers marginal improvement, while the standard Hahn and CPMG sequences provide substantially better protection, with CPMG exhibiting superior performance over the other ``pure'' sequences in this noise regime, particularly at longer times.

Crucially, the sequence generated by \sol{} significantly outperforms all standard DD sequences (UDD, Hahn, CPMG) as time progresses. At the maximum evaluated time of $T=200\,\text{\textmu s}$, \sol{}'s learned sequence maintains an average coherence of approximately 0.45, whereas the next best standard sequence, CPMG, decays to roughly 0.1, demonstrating the substantial advantage conferred by the adaptive sequence construction. Further, the performance of \sol{}'s sequence closely tracks that of the Oracle sequence, which represents the theoretical optimum achievable via segment selection with perfect a priori noise knowledge. This indicates that the RL agent successfully discovers near-optimal strategies and that \sol{} scales effectively, maintaining its high relative performance even as the total evolution time increases. This highlights \sol{} 's ability to find robust, high-fidelity control sequences tailored to the specific noise characteristics.


\vspace{2mm}

\begin{figure}[t]
    \centering
    \includegraphics[width=0.99\linewidth]{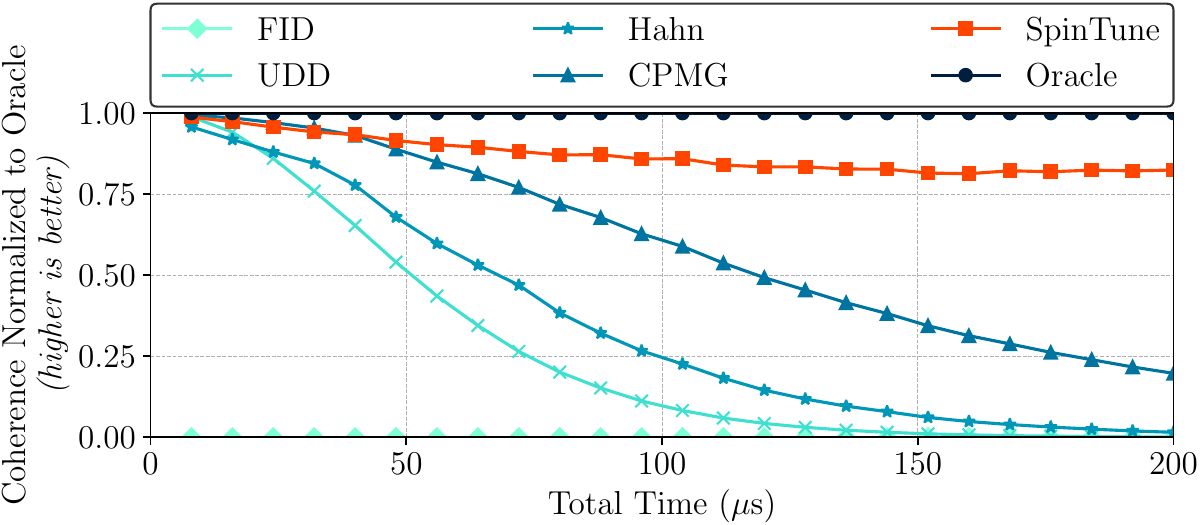}
    \vspace{-3mm}
    \caption{When the coherence is normalized to the Oracle's coherence, we observe that the coherence of \sol{} stabilizes at about 82\% of the Oracle's performance, while baseline DD techniques suffer degradation.}
    \vspace{-5mm}
    \label{fig:norm_coh}
\end{figure}

\noindent\textbf{(III) To further quantify the relative effectiveness of the different strategies, on average, \sol{}'s learned sequence yields a 62\% point higher normalized coherence when compared to the next best sequence tested, CPMG.}  Fig.~\ref{fig:norm_coh} presents the mean coherence of each sequence type normalized by the mean coherence achieved by the Oracle sequence at each corresponding time point. This normalization effectively scales the oracle's performance to 1.0, allowing for a direct assessment of how efficiently each strategy utilizes the available coherence protection potential as the evolution time $T$ increases. The plot illustrates the advantage of the adaptive strategy learned by \sol{}. The \sol{} sequence consistently maintains a remarkably high fraction of the Oracle's performance, starting near 100\% efficiency and degrading only slowly to retain over 82\% of the optimal average coherence even at the evolution time of $T=200\,\text{\textmu s}$.

In contrast, the standard DD sequences exhibit a much more rapid decline in efficiency relative to the Oracle. While CPMG initially approaches Oracle-level performance, its normalized coherence drops significantly, retaining only about 20\% of Oracle's coherence by $T=200\,\text{\textmu s}$. Other standard sequences like UDD and Hahn decay even more rapidly in relative terms, achieving only a small fraction of the optimal coherence at intermediate times and becoming largely ineffective at longer times. The FID sequence, as expected, shows negligible normalized coherence throughout.

\vspace{2mm}

\begin{figure}[t]
    \centering
    \includegraphics[width=0.99\columnwidth]{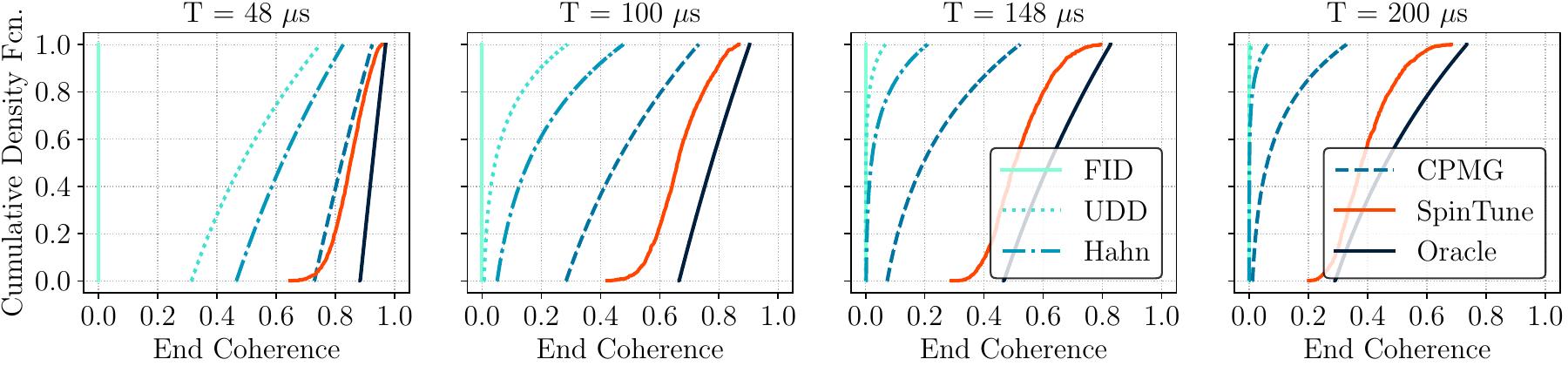}
    \vspace{1mm}
    \vspace{-3.5mm}
    \caption{We show the cumulative density function (CDF) of the coherence achieved with different techniques at four different time steps. Compared to baseline DD pulses, \sol{} consistently achieves coherence that is much higher and close to that of the Oracle technique.}
    \vspace{-5mm}
    \label{fig:main_cdfs}
\end{figure}

\noindent\textbf{(IV) \sol{} consistently achieves near-optimal coherence with low variability across diverse noise environments.} Having assessed the mean performance, we now examine variability across noise environments. Fig.~\ref{fig:main_cdfs} presents the CDFs of coherence for different strategies under all simulated Noise Spectral Density (NSD) settings at various evolution times. Note: $T=48 \, \mu$s and $T=148 \, \mu$s are shown as they represent the closest time steps to $T=50 \, \mu$s and $T=150 \, \mu$s where pulses can end due to their quantized time durations. A curve to the right indicates a higher probability of achieving a small difference from the Oracle, indicating better, more consistent performance. Across all evaluated times, the curve corresponding to \sol{} is distinctly positioned to the right of all standard DD sequences and close to the curve of the Oracle. For instance, at $T=48 \, \mu$s, about 80\% of the noise environments simulated achieve a coherence of between 0.8 and 1.0 with \sol{}. In contrast, the standard DD sequences exhibit significantly more variable performances. CPMG, the best-performing standard sequence on average, still shows a considerably larger spread and median difference compared to \sol{}. At $T=48 \, \mu$s, the median coherence for CPMG is 0.62, meaning half of the simulations resulted in coherence of less than 62\%. This underscores that \sol{} not only achieves higher average coherence but also does so more reliably and consistently across a wide range of noise conditions than non-adaptive DD strategies.

\vspace{2mm}

\begin{figure}[t]
    \centering
    \includegraphics[width=0.99\linewidth]{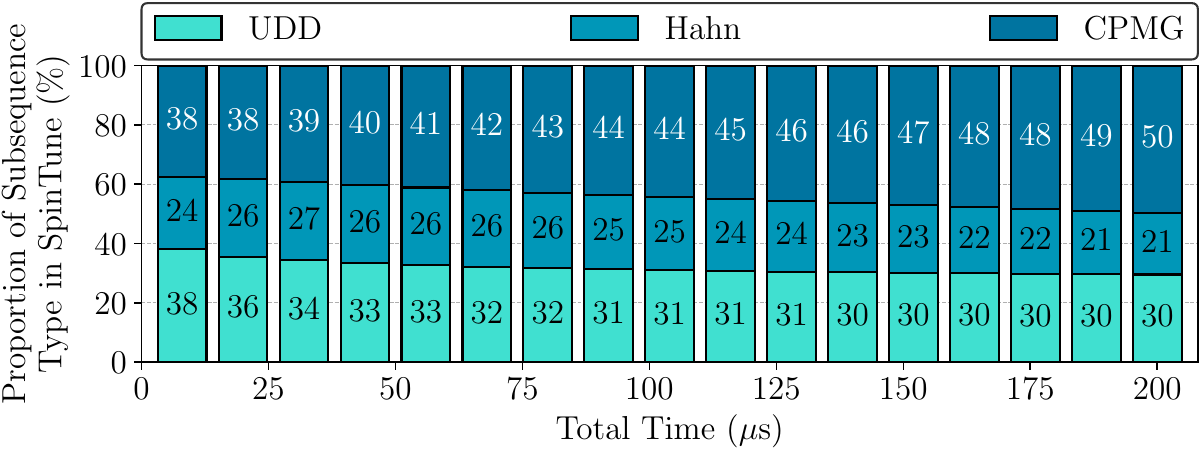}
    \vspace{-3mm}
    \caption{\sol{}'s distribution of different standard subsequences as the total evolution time $T$ increases.}
    \vspace{-5mm}
    \label{fig:prop_seqs}
\end{figure}

\noindent\textbf{(V) UDD, Hahn, and CPMG subsequences, all make substantial contributions to \sol{}'s sequences, and the relative contribution of each subsequence evolves over time.} To understand the strategy employed by \sol{}'s reinforcement learning agent that leads to the superior coherence preservation observed, we analyze the composition of the learned sequences. Fig.\ref{fig:prop_seqs} illustrates the average proportion of each fundamental DD subsequence type within the learned sequences as a function of the total evolution time $T$. The most striking observation is \sol{}'s utilization of a dynamic mixture of different DD subsequences, contrasting with the pure sequence strategies, which, by definition, employ only a single type. The FID subsequence, which offers no noise protection, is entirely absent from the learned solutions, indicating the agent effectively learns to avoid detrimental actions.

Instead, \sol{}'s sequences are consistently composed of UDD, Hahn, and CPMG subsequences. Even though UDD and Hahn perform particularly poorly on their own, \sol{}'s careful placement of them helps improve the overall coherence. The relative contribution of these types evolves with the total sequence duration $T$. At shorter evolution times, Hahn echoes constitute a substantial fraction (around 24\%), alongside significant contributions from both CPMG and UDD. As the total time $T$ increases, the proportion of Hahn and UDD subsequences gradually decreases; concurrently, the proportion of CPMG subsequences steadily rises, becoming a 50\% contributor at longer times. This adaptive composition suggests that the performance advantage of the \sol{} stems from its ability to flexibly combine the different filtering characteristics of UDD, Hahn, and CPMG subsequences. By dynamically adjusting the mixture based on the total evolution time, \sol{} can construct a composite sequence whose overall filter function is better tailored to suppress specific noise environments compared to any sequence restricted to a single DD type. 

\vspace{2mm}

\begin{figure}[t]
    \centering
    \includegraphics[width=0.99\linewidth]{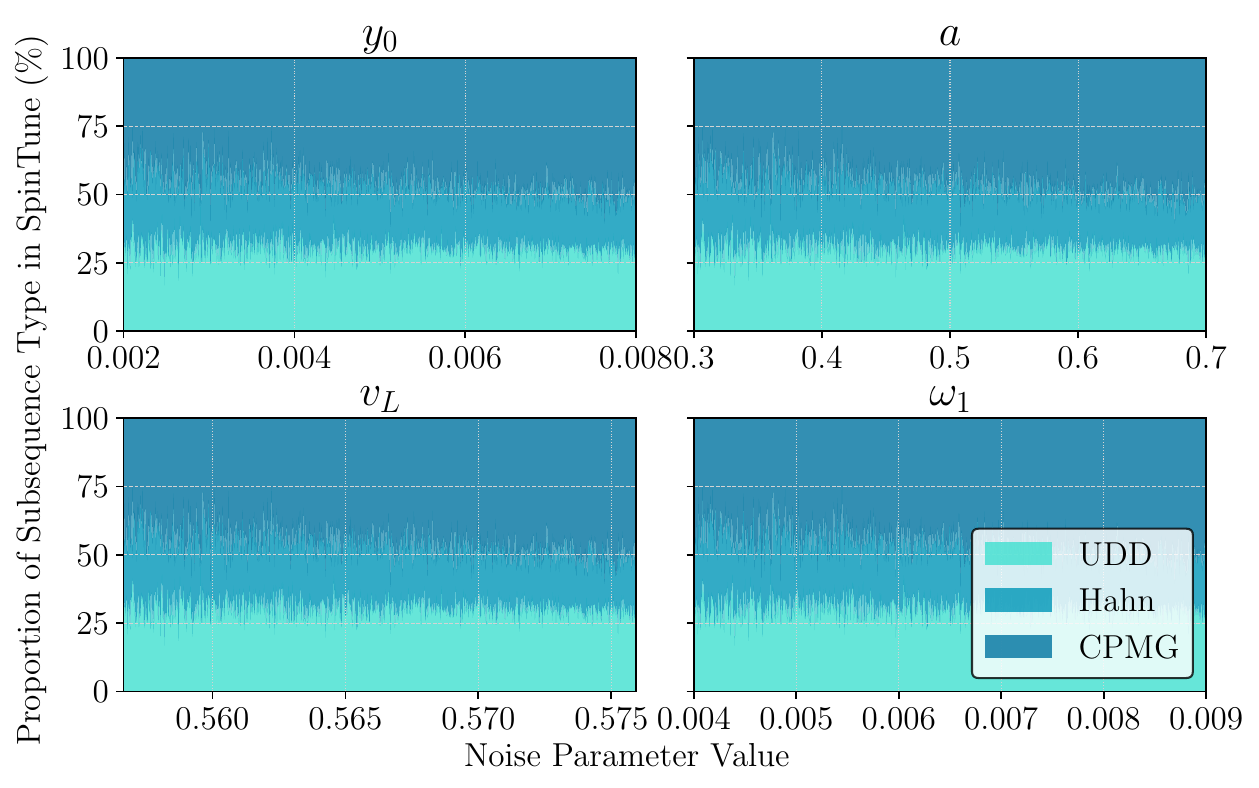}
    \vspace{-3mm}
    \caption{\sol{}'s distribution of different standard subsequences as the values of different noise parameters vary.}
    \vspace{-5mm}
    \label{fig:noise_pars}
\end{figure}

\noindent\textbf{(VI) \sol{} adapts to different noise profiles by varying the subsequence mixture.} To investigate how \sol{} adapts its strategy to different noise characteristics, Fig.~\ref{fig:noise_pars} examines the average composition of \sol{}'s sequences as each individual noise parameter ($y_0$, $a$, $v_L$, $w_1$) is varied across its simulated range. Recall that the noise Gaussian is modeled as: $S(\omega) = y_0 + a \exp\left(-\frac{(\omega-v_L)^{2}}{2w_1^{2}}\right
)$. Each subplot displays the proportions of UDD, Hahn, and CPMG subsequences, averaged over all evolution times, for runs corresponding to the specific noise parameter value on the x-axis.

The most prominent feature across all four subplots is the relative stability of the subsequence distribution. \sol{}'s RL agent consistently employs a mixture primarily composed of all UDD, Hahn, and CPMG subsequences, with no use of FID, regardless of the specific parameter values. This indicates that the general strategy of utilizing a diverse set of DD building blocks is robust across the simulated variations in noise offset ($y_0$), amplitude ($a$), center frequency (related to $v_L$), and width ($w_1$). CPMG and UDD consistently represent the largest fractions, with Hahn contributing a smaller but significant portion. While significant shifts in strategy are not observed based on these individual parameter variations, some subtle trends can be discerned. As the parameters increase in magnitude, there appears to be a tendency for the agent to decrease the usage of Hahn subsequences while slightly increasing the proportion of CPMG subsequences.

\vspace{2mm}

\begin{figure}[t]
    \centering
    \includegraphics[width=0.99\linewidth]{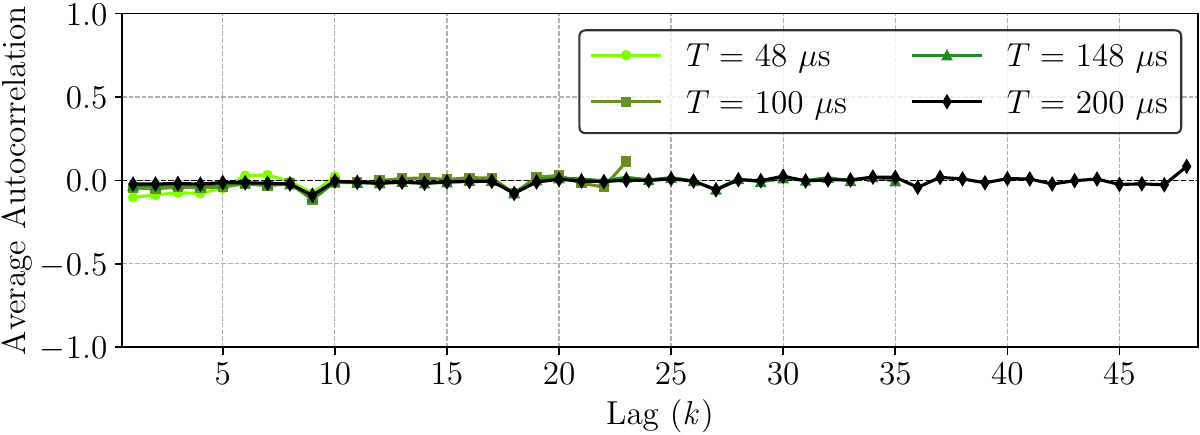}
    \vspace{-3mm}
    \caption{The sequences generated by \sol{} exhibit a low autocorrelation at all lags, indicating non-repetitive patterns.}
    \vspace{-5mm}
    \label{fig:auto_corr}
\end{figure}

\noindent\textbf{(VII) Learned sequences exhibit low temporal correlation, indicating complex, non-repetitive structures.} To further probe the structure of the sequences generated by \sol{}, we analyze the autocorrelation of the chosen subsequence types. Fig.~\ref{fig:auto_corr} displays the average autocorrelation for the sequence of subsequence types as a function of lag $k$ (separation between subsequence positions) for different evolution times. The autocorrelation measures the Pearson correlation~\cite{cohen2009pearson} between subsequences at positions $p$ and $p+k$ averaged over the entire sequence and all simulations.

The plot shows that, for all evaluated evolution times $T$, the average autocorrelation is close to zero for lags $k > 0$, indicating very low correlation. While there are minor fluctuations, no significant positive or negative correlation persists across different lags or evolution times. This low autocorrelation indicates the complexity of the learned strategies. It demonstrates that the sequences generated by \sol{} do not rely on simple, repetitive patterns of a single subsequence type (e.g., repeating CPMG blocks). Instead, the RL agent successfully learns to construct more intricate, aperiodic sequences where the selection of subsequences is not naively temporally correlated -- some examples are shown in Fig.~\ref{fig:example_seq}. This avoidance of simple repetition highlights the non-trivial nature of the control problem and underscores \sol{}'s ability to discover sophisticated, dynamically varying control protocols rather than converging to easily learnable periodic solutions.

%% file: sections/casestudy.tex
\section{Case Study: Applying \sol{} to a Neutral Atom Quantum Sensing System Simulated on a Neutral Atom Quantum Computer}
\label{sec:case_study_neutral_atom}

To validate the adaptability and effectiveness of our framework beyond the simulated noise environments of NV centers, we applied \sol{} to a physical quantum computer. While we mainly discussed SpinTune in the context of NV centers, as it is difficult to access an NV center testbed for testing, we instead ran a validation experiment on a neutral atom computer, which has similar physical dynamics at a high level. We accessed a commercially available analog neutral-atom platform, QuEra's Aquila device, via Amazon Braket. While the neutral-atom setting is fundamentally different from the NV-center setting, we perform these experiments to demonstrate that \sol{} can characterize the noise of an arbitrary quantum architecture and generate a DD sequence to enhance its coherence. The process involved three key stages: (1) empirical hardware noise characterization, (2) noise model reconstruction, and (3) \sol{}-driven sequence optimization and execution.

\vspace{2mm}
\noindent\textbf{Methods.} The first step was to construct a characterization of the noise processes affecting the qubits on the Aquila hardware. Unlike the NV center NSD, where the noise source is understood as a $^{13}\text{C}$ spin bath, the noise in a neutral atom system arises from a combination of factors like laser fluctuations, magnetic field effects, and atomic motion~\cite{wurtz2023aquilaquera}. Therefore, we opted to design an empirical noise model that captures Aquila's behavior, which we could use to generate a neutral-atom NSD. We thus performed two sets of experiments on a noisy simulation of the Aquila hardware via the Bloqade library to probe its decoherence properties under different conditions~\cite{bloqade2023quera}. We opted to use a noisy Aquila model for noise spectroscopy rather than real hardware runs, since we needed a large number of shots (~100K). The first set of shots was a Ramsey experiment, which measures coherence during free evolution (no additional pulses). The second set was a CPMG sequence with 8 equidistant pulses. For robustness, we ran these experiments across a range of evolution times and multiple samples at each to obtain reliable measurements of the qubit's final-state probabilities, which serve as sensor readings. With this empirical data, we then reconstructed the underlying NSD of the hardware. We proposed a three-component model for the NSD:
\begin{equation}
\textstyle     S(\omega) = y_0 + a_g \exp\left(-\frac{(\omega - v_g)^2}{2w_g^2}\right) + \frac{a_{1f}}{\omega}
    \label{eq:neutral_atom_nsd}
\end{equation}
This model includes a constant white noise floor $y_0$, a Gaussian with amplitude $a_g$, center frequency $v_g$, and width $w_g$, along with a $1/f$ term with amplitude $a_{1f}$.

\begin{figure}[t]
    \centering
    \includegraphics[width=0.99\linewidth]{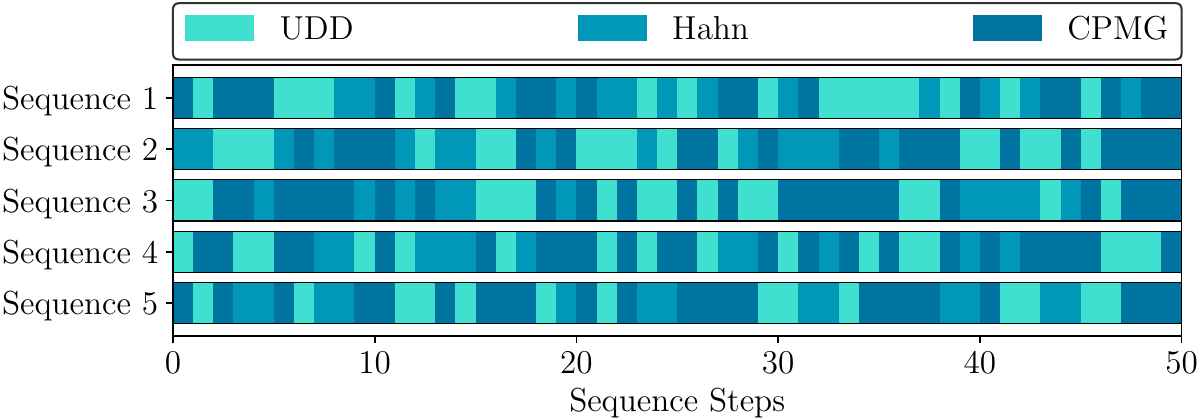}
    \vspace{-3mm}
    \caption{Examples of sequences generated by \sol{}'s RL engine highlight their variety and complexity.}
    \vspace{-5mm}
    \label{fig:example_seq}
\end{figure}

\begin{figure*}[t]
    \centering
    \subfloat[\label{fig:model_validation}]
    {\includegraphics[width=0.49\linewidth]{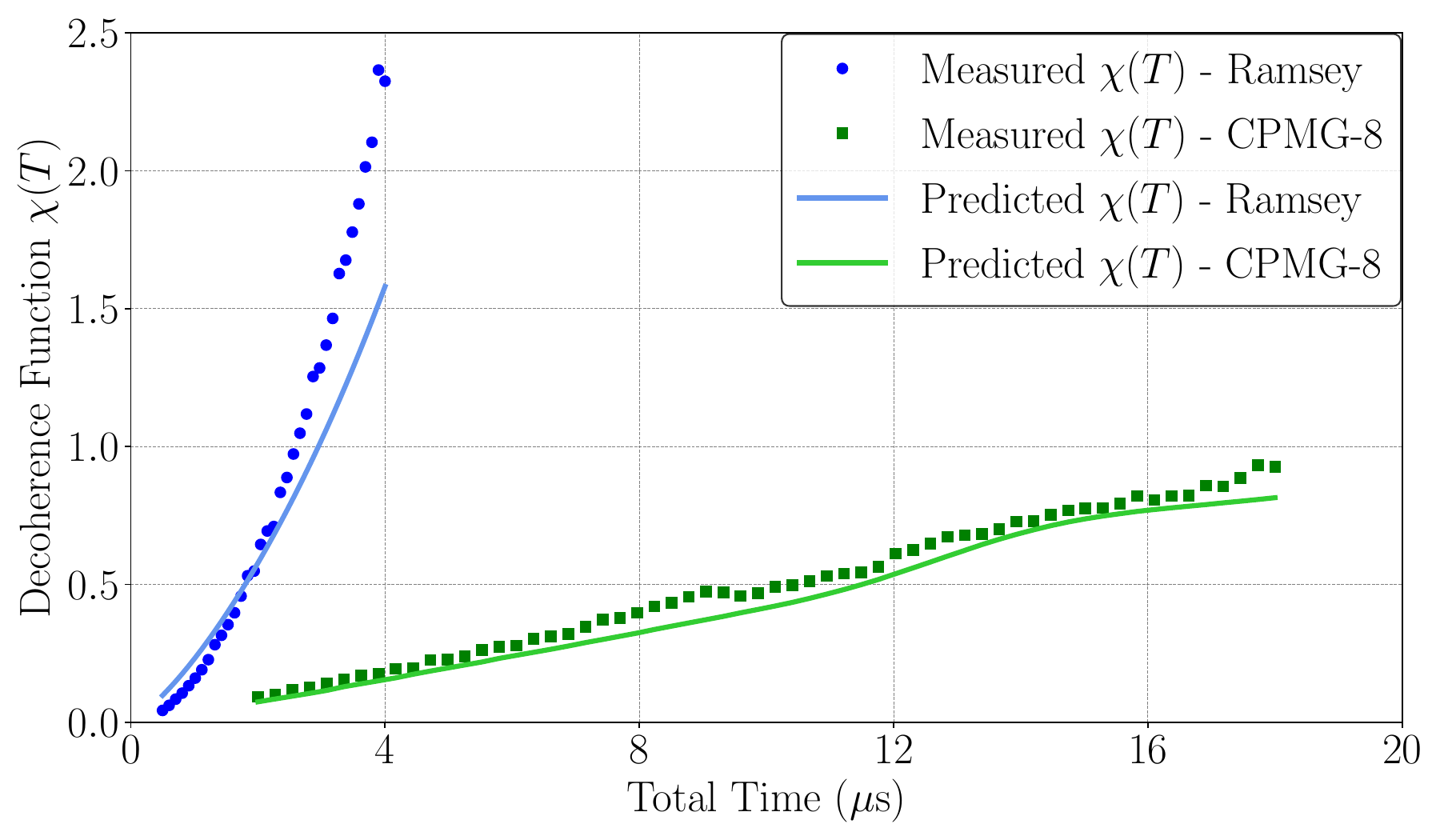}}
    \hfill 
    \subfloat[\label{fig:hardware_results}]
    {\includegraphics[width=0.51\linewidth]{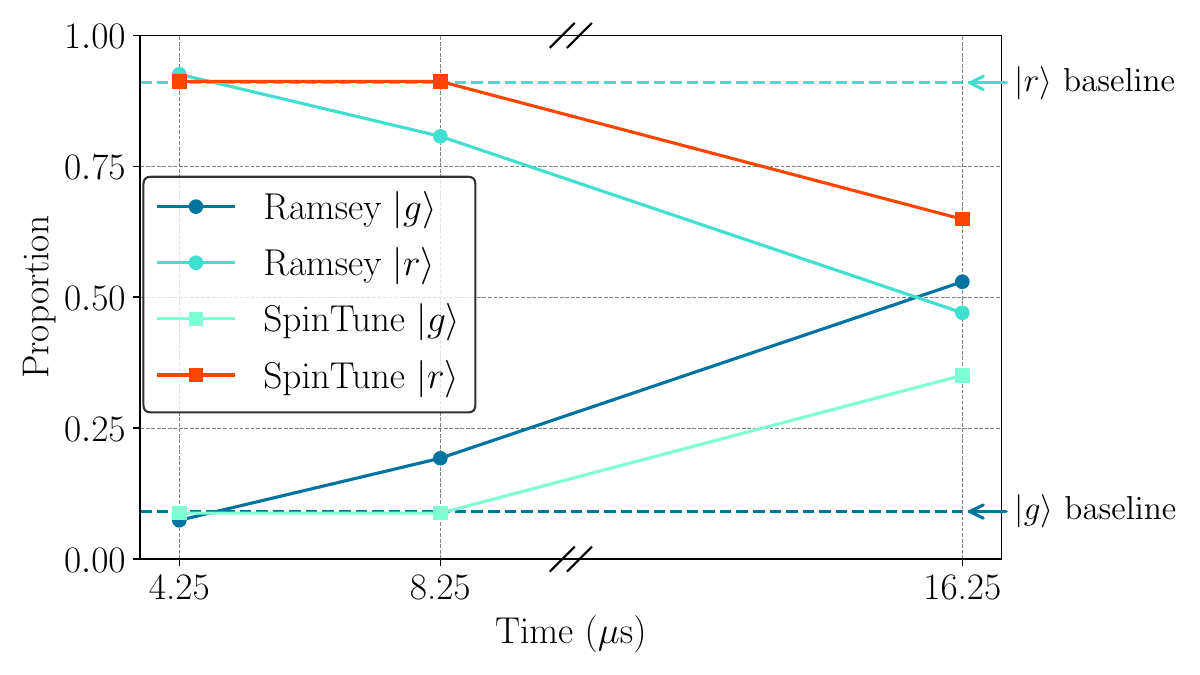}}
    \label{fig:case_study_combined}
    \vspace{-3mm}
    \caption{\textbf{(a)} Validation of the fitted noise model against empirical data from simulated noisy runs of Aquila. The points represent decoherence data from the simulations; the solid lines represent the predicted decoherence from our fitted NSD. \textbf{(b)} Results from Aquila comparing the performance of the \sol{}-generated sequence against a Ramsey free-evolution sequence. The lines show the proportions of ground ($|g\rangle$) and Rydberg ($|r\rangle$) states at different evolution times, with higher $|r\rangle$ proportions indicating greater coherence. The dashed horizontal lines represent the baseline: a $0.25~\mu\text{s}$ Ramsey evolution.}
    \vspace{-4mm}
\end{figure*}

We performed a fitting procedure that simultaneously optimized the parameters of Eq.~\ref{eq:neutral_atom_nsd} to match the measured decoherence data from both the Ramsey and CPMG experiments. The fit was highly successful, which is shown in Figure~\ref{fig:model_validation}.

Having derived and validated a mathematical model of the Aquila device's noise, we swapped the new neutral-atom NSD function for \sol{}'s previous NV-center-based function. We then ran \sol{} for a total evolution time of $16~\mu\text{s}$, with the reward function being the final coherence. The RL agent converged on a sequence constructed from standard DD blocks. For the target duration of $16~\mu\text{s}$, the optimal strategy was found to be the four-component sequence UDD $\Rightarrow$ CPMG $\Rightarrow$ CPMG $\Rightarrow$ CPMG. Finally, we translated this \sol{}-generated sequence into the pulse-level control language of the Aquila hardware using the Amazon Braket SDK. We then deployed this optimized sequence, along with a baseline Ramsey evolution sequence for comparison, for execution on the physical Aquila computer.

Both the \sol{} sequence and the Ramsey sequence were run on the Aquila device for progressively longer evolution times: $4.0~\mu\text{s}$, $8.0~\mu\text{s}$, and $16.0~\mu\text{s}$. An additional $0.25~\mu\text{s}$ was added to each total time to account for the duration of the initial and final $\pi/2$ pulses. The results of these hardware experiments are presented in Figure~\ref{fig:hardware_results}. The plot shows the proportion of qubits measured in the ground ($|g\rangle$) state versus the excited Rydberg ($|r\rangle$) state for both sequences. Since the experiments are designed to measure the preservation of a quantum superposition state, the goal is to maximize the proportion of qubits measured in the Rydberg state. A departure from pure Rydberg towards a 50/50 mixture of $|g\rangle$ and $|r\rangle$ indicates a loss of coherence. 

\vspace{2mm}

\noindent\textbf{Results.} As expected, as the evolution time increases, decoherence increases with both sequences. However, a significant performance gap emerges as the evolution time increases. At $8.25~\mu\text{s}$, the Ramsey sequence already shows notable decay, while the \sol{} sequence maintains a higher proportion of coherent qubits measured in the Rydberg state. At $16.25~\mu\text{s}$, the Ramsey sequence has completely decohered, with the measured state proportions reaching the 50/50 distribution of a fully mixed state where all quantum information has been lost. In contrast, \sol{}'s sequence successfully mitigates the hardware noise, with approximately 66\% of qubits remaining in the $|r\rangle$ state.

This on-hardware comparison provides compelling evidence of \sol{}'s success. By first empirically modeling Aquila's noise profile and then using that model to guide \sol{}, we were able to autonomously generate a custom DD sequence that substantially extends qubit coherence on a physical quantum computer. This result validates \sol{} as a practical method for enhancing the performance of real-world quantum systems.

%% file: sections/related_work.tex
\section{Related Work}
\label{sec:related_work}

In the context of NV centers and DD sequences, our work builds on several foundational efforts in quantum control and sensing. Traditional DD protocols such as Hahn echo, CPMG, and UDD were developed in the NMR and quantum information literature to suppress environmental noise through analytically designed pulse patterns~\cite{hahn1950spin,meiboom1958modified,uhrig2007keeping}. These sequences are constructed to cancel specific orders of decoherence using symmetry or filter design principles, but are not generally optimal across diverse noise profiles. To improve robustness, composite sequences like XY-family protocols apply pulses along alternating axes with the goal of mitigating hardware-specific pulse errors~\cite{souza2011robust}, while concatenated DD sequences hierarchically embed sub-sequences to suppress both low- and high-frequency components~\cite{khodjasteh2009dynamically}. While these hand-crafted strategies have substantially extended NV coherence times, they are still limited by their inability to adapt to arbitrary or unknown spectral noise environments.

To address this, more recent work has explored automated sequence design using algorithmic search and machine learning. One direction uses black-box optimization to tune continuous control waveforms, achieving longer coherence times than textbook sequences by directly optimizing over the control landscape~\cite{miao2022}. Another class of methods employs noise spectroscopy: by measuring NV coherence under a range of known DD sequences, these approaches reconstruct the noise spectral density~\cite{hernandez18,Martina_2023}. While these works focus on learning the noise environment of an NV Center, \sol{} instead uses a reinforcement learning agent to implicitly learn which sequences best suppress decoherence by interacting with the environment through simulated feedback.

Orthogonal to control-sequence optimization, quantum sensing with Rydberg-atom receivers has recently shown an order-of-magnitude finer granularity and end-to-end applications in a wireless setting~\cite{zhang_sensing2023,jiao2024sensing}. Our work can be combined with this work to further enhance quantum sensing.

RL has also recently begun to see applications in quantum computing control. Bukov~\textit{et al.}~\cite{bukov2018} showed that RL agents could learn pulse sequences to drive quantum systems toward target states, and related work in adaptive experimental design applies Bayesian inference to adjust parameters in real time based on measurement feedback~\cite{wang2022}. However, such online learning is infeasible at the timescales relevant to NV magnetometry. In contrast, \sol{} performs offline training and outputs a single high-fidelity DD sequence, thereby enabling scalable learning without runtime overhead and making it practical for quantum sensing. Moreover, while many established methods, ranging from traditional DD protocols to optimized approaches using genetic algorithms~\cite{quiroz2013,farfurnik15,Hall2010}, also succeed in reducing coherence loss, they typically require intensive, noise-specific re-optimization to work effectively under varying conditions. \sol{} not only generalizes well for arbitrary noise due to its offline RL training, but is also optimized computationally.

%% file: sections/conclusion.tex
\section{Conclusion}
\label{sec:conclusion}

In this work, we introduced \sol{}, an RL-guided software framework that autonomously discovers high-fidelity dynamical decoupling sequences tailored to suppress decoherence in NV-center and neutral-atom-based quantum sensors. By leveraging filter-function-based coherence evaluation with memoized Fourier transforms, \sol{} efficiently learns adaptive pulse patterns across different noise environments. Our extensive evaluations demonstrate that \sol{} consistently outperforms analytically derived baselines and approaches theoretical Oracle performance, extending coherence and significantly improving quantum sensing sensitivity. Beyond advancing quantum control, \sol{} provides a scalable, deployable method that improves the reliability and robustness of quantum sensors. \textit{This positions \sol{} as an enabling layer for integrating quantum sensors into future quantum-classical HPC systems, making quantum-enhanced sensing practical for broader scientific and machine learning applications.}

\section*{Acknowledgement} 

Grammarly AI assistant was used to edit this work, and all generated text was verified by the authors. This work was supported by Rice University, the Rice University George R. Brown School of Engineering and Computing, and the Rice University Department of Computer Science. This work was supported by the Ken Kennedy Institute and the Rice Quantum Initiative, which is part of the Smalley-Curl Institute.